\documentclass[aps,prb,reprint,superscriptaddress,secnumarabic,amssymb]{revtex4-1}
\usepackage{graphicx}
\usepackage{dcolumn}
\usepackage{bm}
\usepackage{color}
\usepackage{bbold}
\usepackage[breaklinks=true,colorlinks,citecolor=blue,linkcolor=blue,urlcolor=blue]{hyperref}

\newcommand{\fref}[1]{Fig.\hspace{0.025in}\ref{#1}}
\newcommand{\sref}[1]{Sec. \ref{#1}}
\newcommand{\aref}[1]{App. \ref{#1}}
\newcommand{\eref}[1]{Eq.\hspace{0.025in}(\ref{#1})}

\usepackage{graphicx}
\usepackage[normalem]{ulem}
\usepackage{color}
\usepackage{amsmath}
\usepackage{enumitem}

\newcommand\wordcount{
    \immediate\write18{texcount -sub=section \jobname.tex  | grep "Section" | sed -e 's/+.*//' | sed -n \thesection p > 'count.txt'}
(\input{count.txt}words)}

\begin{document}
\title{Quantum dissipation with conditional wave functions: \\Application to the realistic simulation of nanoscale electron devices}

\author{E. Colom\'es}
\affiliation{Departament d'Enginyeria Electr\`onica, Universitat Aut\`onoma de Barcelona, 08193-Bellaterra (Barcelona), Spain}
\author{Z. Zhan}
\affiliation{Departament d'Enginyeria Electr\`onica, Universitat Aut\`onoma de Barcelona, 08193-Bellaterra (Barcelona), Spain}
\author{D. Marian}
\affiliation{Dipartimento di Ingegneria dell'Informazione, Universit\`a di Pisa. Via G. Caruso 16, 56122 Pisa, Italy.}
\author{X. Oriols}
\email{xavier.oriols@uab.cat}
\affiliation{Departament d'Enginyeria Electr\`onica, Universitat Aut\`onoma de Barcelona, 08193-Bellaterra (Barcelona), Spain}

\begin{abstract}
Without access to the full quantum state, modeling dissipation in an open system requires approximations. The physical soundness of such approximations relies on using realistic microscopic  models of dissipation that satisfy completely positive dynamical maps. Here we present an approach based on the use of the Bohmian conditional wave function that, by construction, ensures a completely positive dynamical map for either Markovian or non-Markovian scenarios, while allowing the implementation of realistic dissipation sources. Our approach is applied to compute the current-voltage characteristic of a resonant tunneling device with a parabolic-band structure, including electron-lattice interactions. A stochastic Schr\"odinger equation is solved for the conditional wave function of each simulated electron. We also extend our approach to (graphene-like) materials with a linear band-structure using Bohmian conditional spinors for a stochastic Dirac equation.
\end{abstract}

\maketitle


\section{Introduction}
\label{s1}
Although reversible dynamics in a closed system induce irreversibility into a smaller subsystem, the simulation of quantum dissipation cannot rely on the full quantum state, because it is computationally inaccessible. The solution is to deal only with the degrees of freedom of a smaller subsystem, referred as the open system \cite{open}, or simply the system. The remaining degrees of freedom constitute the environment. Most approaches for open systems revolve around the reduced density matrix constructed by tracing out the degrees of freedom of the environment \cite{open}. A proper equation of motion of the reduced density matrix must lead to a dynamical map that satisfies complete positivity (CP) \cite{positivity}, which guarantees that such a reduced density matrix is always a positive operator. Some phenomenological treatments of the source of dissipation violate CP, such as the Boltzmann collision operator in the Liouiville equation \cite{Zhen} or the seminal Caldeira Leggett master equation \cite{caldeira}. For Markovian evolutions, the Lindblad master equation \cite{lindblad} preserves CP, but its connection to realistic practical scenarios and its extension beyond Markovian dynamics are still challenging \cite{ferialdi,vega}.

Alternatively, inspired by the spontaneous collapse theories~\cite{GRW}, Di\'osi, Gisin, and Strunz developed the stochastic Schr\"odinger equations (SSEs) to \emph{unravel} the reduced density matrix in non-Markovian systems \cite{SSE}. Continuous measurement theory allows the definition of a wave function of the open system conditioned on one \emph{monitored} value associated with the environment\cite{wiseman,Gambetta,dioosi}. This approach preserves positivity because the reduced density matrix is built from a sum of projectors associated with the states solution of a Schr\"odinger-like equation\cite{open,vega,Gambetta}. In practical applications, the non-hermitian Hamiltonians can provoke states of the SSE to lose their norm and therefore their statistical relevance\cite{open}. 

Here a discussion about the physical interpretation of the pure-state solution of the SSE is relevant. It is well recognized that the continuous measurement of an open system with Markovian dynamics can be described by a SSE\cite{open}. Therefore, the pure-state solution of SSE can be interpreted as the state of the Markovian system while the environment is under (continuous) observation. However, such a physical interpretation cannot be given to the solutions of the SSE for non-Markovian systems\cite{wiseman,Gambetta,dioosi}. In such non-Markovian systems, a continuous measurement requires a non-trivial interaction of the system with the environment so that the physical description of the continuously measured open system needs to be done through the reduced density matrix \cite{open}(not through the pure-state given by the non-Markovian SSE, which becomes just a numerical tool). The physical interpretation that one can assign to the solution of the non-Markovian SSE (conditioned to some environment value) is the following: the state of the open system at a given time if a measurement is performed in the environment at that time, yielding the mentioned value for the environment. Linking SSE states of the open system (or values of the environment) at different times is just a fiction.

In this work, we present an approach to deal with quantum dissipation based on the use of Bohmian conditional wave functions (CWFs) \cite{Bohm}. Such a CWF provides an unproblematic way of defining the wave function of a subsystem, either from a computational and an interpretative points of view. By construction, within Bohmian mechanics, the CWF is always a well-defined physical state for Markovian and non-Markovian open systems, with continuous or non-continuous measurements. The general expression of the equations of motion of such a CWF with or without dissipation is mentioned in Ref. \onlinecite{PRLxavier}. We anticipate the two main results of this work. First, since our approach deals directly with wave functions, it provides a CP map for either Markovian or non-Markovian dynamics with an unproblematic physical interpretation of the wave function of the open system at different times. Second, contrary to other CP methods, the numerical inclusion of different dissipative phenomena in the equation of motion of the CWF can be done straightforwardly with a  microscopic and realistic implementation. These properties make the approach presented in this work very relevant for many different research fields. In this paper, we discuss its implementation for quantum transport with dissipation.  

The article is structured as follows. After this introduction, in \sref{s2}, we present the basic elements of our general approach. In particular, we discuss its complete positivity in \sref{s2.1}, the equation of motion of the CWF with dissipation in \sref{s2.2} and a comparison with similar techniques in \sref{s2.3}. In \sref{s3}, as an example, we study quantum dissipation through electron phonon interaction, with the definition of the conditional potential in \sref{s3.1}. The application to tunneling nanodevices with parabolic and linear band structures is done in \sref{s3.2}  and \sref{s3.3}, respectively. We conclude in \sref{s6}. Finally, technical details are discussed in four appendices.\\

\section{The approach}
\label{s2}

We consider an \emph{isolated} (\emph{closed}) quantum system described by a full many-body state $|\Psi \rangle$ solution of the unitary, reversible, and linear Schr\"odinger equation. We decompose the total Hilbert space of $N$ particles as $\mathcal{\hat H}=\mathcal{\hat H}_a \otimes \mathcal{\hat H}_b$, with $\mathbb{\vec r} =\{\mathbb{\vec r}_a,\mathbb{\vec r}_b\}$ being $\vec {\mathbb r}_a$  the position of the $a$-particle and $\mathbb{\vec r}_b=\{\mathbb{\vec r}_1,..,\mathbb{\vec r}_{a-1},\mathbb{\vec r}_{a+1},.., \mathbb{\vec r}_N\}$ the position of all other particles. Next, we present our approach emphasizing that it provides a CP map for either Markovian or non-Markovian systems.    

\subsection{Complete positivity}
\label{s2.1}

The expectation value $\langle  O_a \rangle \equiv \langle \Psi | \mathcal{\hat O}_a \otimes  \mathbb{1}_b |\Psi \rangle$ associated with an operator $\mathcal{ \hat O}_a$ acting on the $a$-particle, with $\mathbb{1}_b$ being the identity operator for $\mathcal{\hat H}_b$, can be computed as:
\begin{eqnarray}
\label{avalue}
\langle  O_a \rangle &=& \int d \mathbb{\vec r}_a  O_a\rho(\mathbb{\vec r}_{a},\mathbb{\vec {r'}}_a,t)|_{\mathbb{\vec {r'}}_a=\mathbb{\vec r}_a}
\end{eqnarray}
where $\rho(\mathbb{\vec r}_{a},\mathbb{\vec {r'}}_a,t)$ is the reduced density matrix:

\begin{eqnarray}
\rho(\mathbb{\vec r}_{a},\mathbb{\vec {r'}}_a,t)&=&\int d \mathbb{\vec r}_b \Psi^*(\mathbb{\vec {r'}}_a,\mathbb{\vec r}_{b},t)\Psi(\mathbb{\vec r}_a,\mathbb{\vec r}_{b},t),
\label{adensity}
\end{eqnarray}
where $\Psi(\mathbb{\vec r}_a,\mathbb{\vec r}_b,t)\equiv\langle \mathbb{\vec r}_a,\mathbb{\vec r}_b|\Psi \rangle$ is the total wave function and $O_a$ is the position representation of $\mathcal{\hat O}_a$.

The same system can be described with the Bohmian theory\cite{Bohm,ABM} as follows. For each experiment, labeled by $j$, a Bohmian quantum state is defined by the same wave function $\Psi(\mathbb{\vec r}_a,\mathbb{\vec r}_b,t)$ plus a set of well-defined trajectories in physical space, $\{\mathbb{\vec r}_1^j[t],..., \mathbb{\vec r}_N^j[t]\}$. The velocity of each trajectory is \cite{Bohm}:
\begin{equation}
\label{velo}
\vec v^j_a[t]=\frac{d \mathbb{\vec r}^j_a[t]}{dt}=\frac {\vec J_a(\mathbb{\vec r}_a^j[t],\mathbb{\vec r}_b^j[t],t)}{|\Psi(\mathbb{\vec r}_a^j[t],\mathbb{\vec r}_b^j[t],t)|^2} ,
\end{equation}
where $\vec J_ a = \hbar \mathop{\rm Im} (\Psi^* \nabla_a \Psi)/m_a$ is the (ensemble value of the) current density with $m_a$ the mass of the $a$-th particle. The set of $N$ positions $\{\mathbb{\vec r}_1^j[t],..., \mathbb{\vec r}_N^j[t]\}$ in different $j=1,...,W$ experiments is distributed (in quantum equilibrium \cite{Bohm}) as: 
\begin{eqnarray}
\label{QE}
|\Psi(\mathbb{\vec r}_a,\mathbb{\vec r}_b,t)|^2 =  \frac{1}{W} \sum_{j=1}^{W} \delta(\mathbb{\vec r}_a-\mathbb{\vec r}_a^j[t]) \delta(\mathbb{\vec r}_b-\mathbb{\vec r}_b^j[t]).
\end{eqnarray}
The identity in \eref{QE} requires $W \to \infty$. Numerically, we just require a large enough $W$. 

The key element of our approach is the CWF associated with the $a$-th particle in the open system during the $j$-th experiment, defined as $\psi_a^j(\mathbb{\vec r}_a,t)\equiv \Psi(\mathbb{\vec r}_a,\mathbb{\vec r}_b^j[t],t)$. We emphasize that $\psi_a^j(\mathbb{\vec r}_a,t)$ provides an unproblematic (Bohmian) definition of the wave function of an open system\cite{Bohm}. We compute a different CWF for each simulated particles of the open system and for each simulated experiment. In \sref{s2.2} we will discuss the equation of motion of such CWFs.  

Next, we construct the reduced density matrix, \eref{adensity}, using the fundamental elements of the Bohmian theory to shows that our approach based on CWFs is CP.  We define the (tilde) CWF of the $a$-th particle in the $j$-th experiment as:
\begin{eqnarray}
\label{tilde}
\tilde \psi_a^j(\mathbb{\vec r}_a,t) \equiv \frac{\Psi(\mathbb{\vec r}_a,\mathbb{\vec r}^j_{b}[t],t)}{\Psi(\mathbb{\vec r}_a^j[t],\mathbb{\vec r}_b^j[t],t)}
\end{eqnarray}
Notice that the denominator $\Psi(\mathbb{\vec r}_a^j[t],\mathbb{\vec r}_b^j[t],t)$ is just a pure time dependent term (without spatial dependence) that has no net effect on the definition of the velocity in \eref{velo}. The Bohmian velocity of the $a$-particle computed from  $\tilde \psi_a^j(\mathbb{\vec r}_a,t)$ is exactly the same value as the one we get from $\psi_a^j(\mathbb{\vec r}_a,t)$. Putting \eref{QE} into $\langle  O_a \rangle \equiv \langle \Psi | \mathcal{\hat O}_a \otimes  \mathbb{1}_b |\Psi \rangle$, integrating all degrees of freedom and using the definition of the (tilde) CWF in \eref{tilde}, we get: 
\begin{eqnarray}
\label{bvalue}
 \langle  O_a \rangle &=& \sum_{j=1}^{W} \left[ p_j {\tilde \psi^{j*}_a}(\mathbb{\vec r}_a,t)  O_a \tilde \psi^j_a(\mathbb{\vec r}_a,t) \right]_{\mathbb{\vec r}_a=\mathbb{\vec r}_a^j[t]},
\end{eqnarray}
where $p_j=1/W$. \eref{bvalue} allows us to compute $\langle  O_a \rangle$ from \eref{avalue} as:
\begin{eqnarray}
\label{bavalue}
\langle  O_a \rangle &=& \int d \mathbb{\vec r}_a \left[  O_a \sum_{j=1}^{W} p_j \tilde \psi_a^{j*}(\mathbb{\vec {r'}}_a,t) \tilde \psi_a^j(\mathbb{\vec r}_a,t) \right]_{\mathbb{\vec {r'}}_a=\mathbb{\vec r}_a=\mathbb{\vec r}_a^j[t]}
\end{eqnarray}
which directly allows the definition of the following density matrix \cite{durr2}:
\begin{eqnarray}
\rho(\mathbb{\vec {r}}_a,\mathbb{\vec {r'}}_a,t)&=&\sum_{j=1}^{W} p_j \tilde \psi_a^{j*}(\mathbb{\vec {r'}}_a,t) \tilde \psi_a^j(\mathbb{\vec r}_a,t),
\label{bdensity}
\end{eqnarray}
The generalization to CWFs with an arbitrary number of particles is straightforward. The time-evolution of \eref{bdensity} ensures, trivially, that the dynamical map associated with our approach is CP. In the position representation, the density operator $\hat \rho=\sum_{j=1}^{W} p_j |\tilde \psi_a^{j*}(t) \rangle \langle \tilde \psi_a^j(t)|$ gives $\langle \mathbb{\vec r}_{o} | \hat \rho | \mathbb{\vec r}_{o} \rangle=\sum_{j=1}^{W} p_j |\langle \mathbb{\vec r}_{o} | \tilde \psi_a^j(t) \rangle |^2 \ge 0$ at any time\cite{fnt} and at any position  $\mathbb{\vec r}_{o}$. The last step to conclude our CP demonstration is quite simple. If the density matrix in \eref{bdensity} is positive, then the diagonal elements of $\langle \mathbb{\vec r}_{o} | \hat \rho | \mathbb{\vec r}_{o} \rangle$ evaluated only at $\mathbb{\vec r}_{o}=\mathbb{\vec r}_a^j[t]$ and defined as $\langle \mathbb{\vec r}_{o} | \hat \rho | \mathbb{\vec r}_{o} \rangle_{B} \equiv \sum_{j=1}^{W} p_j |\langle \mathbb{\vec r}_{o} | \tilde \psi_a^j(t) \rangle |^2 \delta(\mathbb{\vec r}_{o}-\mathbb{\vec r}_a^j[t])\ge 0$ are, by construction, also positive.

In fact, the term $\langle \mathbb{\vec r}_{o} | \hat \rho | \mathbb{\vec r}_{o} \rangle_{B}$ has a very simple interpretation. For the $j$ experiment, the tilde CWF in \eref{tilde} evaluated at $\mathbb{\vec r}_{o}=\mathbb{\vec r}_a^j[t]$ is $\langle\mathbb{\vec r}_a^j[t] | \tilde \psi_a^j(t)\rangle =\tilde \psi_a^j(\mathbb{\vec r}_a^j[t],t) \equiv \Psi(\mathbb{\vec r}_a^j[t],\mathbb{\vec r}^j_{b}[t],t)/\Psi(\mathbb{\vec r}_a^j[t],\mathbb{\vec r}_b^j[t],t)=1$. Then, since $p_j=1/W$, we get $\langle \mathbb{\vec r}_{o} | \hat \rho | \mathbb{\vec r}_{o} \rangle_{B} \equiv  \sum_{j=1}^{W_B} 1/W =W_B/W$ where $W_B$ is just the number of experiments where the position of the trajectory $\mathbb{\vec r}_a^j[t]$ coincides with $\mathbb{\vec r}_{o}$. In conclusion, as far as we are dealing with CWF and Bohmian trajectories in the dynamical description of the quantum systems with dissipation, the CP of our approach is always satisfied (the number $W_B$ of trajectories with position $\mathbb{\vec r}_{o}=\mathbb{\vec r}_a^j[t]$ can be zero, but it cannot be negative).

\subsection{The equation of motion for the CWFs}
\label{s2.2}

Here we develop the equation of motion for the CWF, $\psi_a^j \equiv \psi_a^j(\mathbb{\vec r}_a,t)\equiv \Psi(\mathbb{\vec r}_a,\mathbb{\vec r}^j_{b}[t],t)$. As we discussed, the Bohmian velocities obtained from $\tilde \psi_a^j$ and $\psi_a^j$ are identical. It has been shown in Ref. \onlinecite{PRLxavier} that the (non-tilde) CWF can be computed, in general, from the following single-particle Schr\"odinger-like equation in physical space: 
\begin{eqnarray}
\label{Pseu_Sch}
i \hbar\frac {d \langle  \mathbb{\mathbb{\vec r}} | \Psi \rangle }{dt}\Big|_{\mathbb{\vec r}_b^j[t]}\!\!= \langle \mathbb{\vec r} | \mathcal{\hat H} |\Psi \rangle|_{\mathbb{\vec r}_b^j[t]} \Longleftrightarrow i \hbar\frac {d \psi_a^j}{dt}={H_a} \psi_a^j ,
\end{eqnarray}
where $\mathcal{\hat H}$ is the many-body Hamiltonian and its relation to $H_a$ will be explained next. First, we notice that the relation between $i \hbar {d \langle  \mathbb{\mathbb{\vec r}} | \Psi \rangle }/{dt}|_{\mathbb{\vec r}_b^j[t]}$ and $i \hbar {d \psi_a^j}/{dt}$ on the right and left sides of \eref{Pseu_Sch} is the following:
\begin{eqnarray}
&&i \hbar\frac {d \psi_a^j(\mathbb{\vec r}_a,t)}{dt} = i \hbar\frac {d \langle \mathbb{\vec r}_a,{\mathbb{\vec r}_b^j[t]} | \Psi(t) \rangle }{dt}= \nonumber\\
&&=i \hbar\frac {d \langle \mathbb{\mathbb{\vec r}} | \Psi(t) \rangle }{dt}\Big|_{\mathbb{\vec r}_b^j[t]}+i \hbar\sum_{k=1,k\neq a}^{N} \nabla_k  \langle \mathbb{\mathbb{\vec r}} | \Psi(t) \rangle\Big|_{\mathbb{\vec r}_b^j[t]}  \vec v_k^j[t] \nonumber\\
&&=i \hbar\frac {d \langle \mathbb{\mathbb{\vec r}} | \Psi(t) \rangle }{dt}\Big|_{\mathbb{\vec r}_b^j[t]}\!\!+i B_a(\mathbb{\vec r}_a,{\mathbb{\vec r}_b^j[t]},t),
\label{t04}
\end{eqnarray}
with the conditional imaginary potential $iB_a$ defined as:
\begin{eqnarray}
B_a\equiv \hbar\sum_{k=1,k\neq a}^{N} \nabla_k  \langle \mathbb{\mathbb{\vec r}} | \Psi(t) \rangle\Big|_{\mathbb{\vec r}_b^j[t]}  \vec v_k^j[t]
\label{B}
\end{eqnarray}
where  $\vec v_k^j[t]={d \mathbb{\vec r}_k^j[t]}/{dt}$ is the Bohmian velocity  of the $k$ particle given by \eref{velo}.  Second, once we have defined $B_a$, the term $H_a$ on the right hand side of \eref{Pseu_Sch} can be defined as:
\begin{eqnarray}
H_a=\frac {\langle \mathbb{\vec r} | \mathcal{\hat H} |\Psi(t) \rangle|_{\mathbb{\vec r}_b^j[t]}+iB_a} {\psi_a^j}.
\label{H_a}
\end{eqnarray}
In general, \eref{Pseu_Sch} is non-linear because $H_a$ in \eref{H_a} depends on the wave function itself. In addition, the imaginary conditional potential $iB_a$ indicates  that the evolution of the CWF can be non-unitary.  \eref{Pseu_Sch} includes any type of evolution for the  CWF (not only linear and unitary ones). In particular,  \eref{Pseu_Sch} alone allows the description of irreversible dynamics (energy dissipation) in the open system as required in this work. Obviously, the full wave function $\Psi(\mathbb{\vec r}_a,\mathbb{\vec r}_{b},t)$ satisfies unitary and linear dynamics, with conservation of the total energy \cite{ABM}. 

The key computation for the practical application of our approach is the evaluation of $H_a$ in \eref{H_a}, which allows us to determine an equation of motion for each CWF, ensuring the CP of our approach. The calculation of $\langle \mathbb{\vec r} | \mathcal{\hat H} |\Psi(t) \rangle$ before conditioning depends on the full many-body wave function and it requires educated guesses \cite{PRLxavier}. The potential $B_a$, which contains many-body terms but it does not depend directly on $\mathcal{\hat H}$, will be approximated following Ref. \onlinecite{PRLxavier}. Stochasticity is introduced in \eref{Pseu_Sch} through the term  $H_a$ which accounts for the effect of non-simulated degrees of freedom of the environment in each experiment.

\subsection{Comparison with other techniques}
\label{s2.3}

Several techniques use Bohmian trajectories as a mathematical/computational tool to solve some \emph{reduced} equations of motion \cite{applied}. Here, on the contrary, \eref{QE} guarantees empirical equivalence between Bohmian and standard quantum (non-relativistic) results in the whole closed system. This implies not only the correct description of any smaller portion of the closed system, i.e. our open systems, but also empirical equivalence in the measured values \cite{Bohm}. It is important to emphasize that  Gambetta and Wiseman \cite{Gambetta} pointed out that the only physical continuous-in-time interpretation of the wave functions solution of non-Markovian SSEs, i.e. with back-action from the environment to the system, has to be based on the Bohmian theory. In other words, in spite of its mathematical interest as a computational tool, the \emph{improper} sum of wave functions of an open system in \eref{bdensity} has a problematic ontological meaning within standard quantum mechanics, as indicated by D'Espagnat \cite{improper,proper}. On the contrary, the Bohmian theory allows a \emph{proper} definition of a wave function of an open system with or without continuous measurements, for both Markovian and non-Markovian dynamics \cite{Bohm}. We can always interpret (part of) $\mathbb{\vec r}_b^j[t]$ as the pointer of a measuring apparatus. Therefore, the Bohmian CWF $\psi_a^j(\mathbb{\vec r}_a,t)$ can be thought of as the wave function of SSE conditioned to a continuous observation defined by the (part of) $\mathbb{\vec r}_b^j[t]$ as the pointer. 

To the best of our knowledge, we are the first to develop a practical SSE algorithm using CWFs solutions of \eref{Pseu_Sch}. In the Bohmian framework, the ensemble values can be directly computed from the trajectories and not from the CWF. Therefore, the technical problems of SSE due to norm degradation are avoided in our approach. It is a remarkable fact that the velocity of $\mathbb{\vec r}_a^j[t]$ computed from $\psi_a^j(\mathbb{\vec r}_a,t)$ gives the exact same value as if we use $\Psi(\mathbb{\vec r}_a,\mathbb{\vec r}^j_{b}[t],t)$. Thus, the velocity, as seen in \eref{velo}, is totally independent of the norm of the CWF\cite{Bohm}. This explains why \eref{Pseu_Sch} deals with a non-normalized wave function. 

Since we are dealing with a \emph{realistic} definition (i.e. with a clear ontological meaning) of the wave function of an open system, $\psi_a^j(\mathbb{\vec r}_a,t)$, a relevant advantage is that our approach allows a \emph{realistic} description of the stochastic sources of dissipation (beyond the typical environmental noise sources introduced in SEE\cite{open}), while maintaining CP. Below, as an example, we provide the stochastic conditioned potential of \eref{Pseu_Sch}, which tackles the electron-lattice energy dissipation in tunneling devices.

\section{Application to electron-lattice interaction}
\label{s3}

To analyze the electron-lattice interaction, here, we develop the exact expression for \eref{Pseu_Sch} for electrons interacting with the lattice. For that purpose, we consider $N_e$ electrons with positions $\vec  r=\{\vec r_1,...,\vec r_{N_e}\}$ and $N_h$ ions located at $\vec  R=\{\vec R_1,...,\vec R_{N_h}\}$.  Although not explicitly indicated, $N_h$ includes also all additional particles required to deal with a closed system with the many body wave function $\Psi(\mathbb{\vec r}_a,\mathbb{\vec r}_b,t)\equiv\langle \mathbb{\vec r}_a,\mathbb{\vec r}_b|\Psi \rangle$ mentioned in \sref{s2.1}. To simplify the notation,  hereafter, we define $\vec r=\{\vec r_a,\vec z_a\}$ with $\vec z_a=\{\vec r_1,..,\vec r_{a-1},\vec r_{a+1},.., \vec r_{N_e}\}$. These new variables are related to previous ones through $\mathbb{\vec r}_a=\vec r_a$ and $\mathbb{\vec r}_b=\{\vec z_a,\vec R\}$, with $\mathbb{\vec r}=\{\mathbb{\vec r}_a,\mathbb{\vec r}_b\}$.

We compute the evolution of the full wave function $\Psi(\vec r,\vec R,t)=\Psi(\mathbb{\vec r},t)$ under the effect of the full Hamiltonian $\mathcal{\hat H}$ in \eref{Pseu_Sch}. The position representation of the Hamiltonian $\mathcal{\hat H}$ gives:
\begin{equation}
H(\vec r, \vec R)= K_e(\vec r)+ K_h(\vec R)+ V_{ee}(\vec r)+ V_{hh}(\vec R)+ H_{ep}(\vec r, \vec R),
\label{Hto}
\end{equation}
with $ K_e(\vec r)$ the electron kinetic energies, $ K_h(\vec R)$ the nucleus kinetic energies, $ V_{ee}(\vec r)$ the electron-electron interactions, $ V_{hh}(\vec R)$ the nucleus-nucleus interactions, and $H_{ep}(\vec r, \vec R)$ the total electron-lattice interaction.  The last term can be split into $ H_{ep}= H_{ep,\vec R_0}+H_{ep,\vec u}$. The first term, $H_{ep,\vec R_0}$, corresponds to the interaction of the electrons with the fixed (equilibrium) positions of the ions $\vec R_0$. The second one, $H_{ep,\vec u}$, includes the interaction of the electrons with the displacement of the ions, $\vec u=\vec R-\vec R_0=\{\vec u_1,..., \vec u_{N_h}\}$, and it is the only term that prevents the exact separation of the many-particle wave function. Thus, we rewrite \eref{Hto} as:

\begin{equation}
H(\vec r, \vec R)= H_c(\vec r, \vec R) + H_{ep,\vec u}(\vec r, \vec R),
\end{equation}
with $H_c(\vec r, \vec R)=K_e(\vec r)+ K_h(\vec R)+ V_{ee}(\vec r)+ V_{hh}(\vec R)+ H_{ep,\vec R_0}(\vec r, \vec R_0)$. Finally, the computation of $H_a$ in \eref{H_a} just requires the explicit evaluation of the terms: 
\begin{eqnarray}
\label{condi}
\langle \vec r_a,\vec z_a,\vec R | \mathcal{\hat H}_{ep,\vec u} |\Psi(t) \rangle|_{\vec z_a^j[t],\vec R^j[t]},
\end{eqnarray}
and 
\begin{eqnarray}
\label{condi2}
\langle \vec r_a,\vec z_a,\vec R | \mathcal{\hat H}_c |\Psi(t) \rangle|_{\vec z_a^j[t],\vec R^j[t]}
\end{eqnarray}
The relevant interaction of the (conditional) wave packet with the moving lattice, present in $\hat H_{ep,\vec u}$, will be evaluated in \sref{s3.1} in the second-quantization formalism. The less relevant interaction of the (conditional) wave packet with the fixed (equilibrium) lattice due to $H_c$ present in \eref{condi2} is discussed in the \aref{apc}.

\subsection{Electron-phonon stochastic potential}
\label{s3.1}

Assuming a small displacement of the ions ${\vec u_h}=\vec R_h-\vec R_{h,0}$ from their equilibrium positions $\vec R_{h,0}$, the electron-lattice Hamiltonian for small displacements of ions in the position representation can be written as $H_{ep,\vec u}(\vec r,\vec R)=\sum_{e,h} {\vec u_h} {\partial U(\vec r_e-\vec R_h)}/{\partial \vec R_h}|_{\vec R_{h,0}}$. The (second-quantization) electron-lattice Hamiltonian is then: 
\begin{eqnarray}
\mathcal{\hat H}_{ep,\vec u} =\sum_{e,p} g^{\vec {q}_p}_{\vec k_e} \hat c_{\vec{k}_e+\vec{q}_p}^\dag \hat c_{\vec{k}_e} (\hat b_{\vec{q}_p} + \hat b_{-\vec{q}_p}^\dag),
\label{int}
\end{eqnarray}
with $\hat b_{\vec{q}_p}$ and $\hat b_{\vec{q}_p}^\dag$ being the annihilation and creation operators of the atomic vibrational eigenstate $|\vec{q}_p\rangle$. Similarly, $\hat c_{\vec{k}_e}$ and $\hat c_{\vec{k}_e}^\dag$ are the corresponding operators of the (Bloch) eigenstate $\langle \vec{r}_e |\hat c_{\vec{k}_e}^\dag|0\rangle=\langle \vec{r}_e | \vec{k}_e \rangle=\phi_{\vec{k}_e}(\vec{r}_e)$. The coupling constant  $g^{\vec {q}_p}_{\vec k_e}$ specifies the transition between the eigenstates. The first-quantization explanation of the electron-lattice interaction and the definition of $g^{\vec {q}_p}_{\vec k_e}$ are given in \aref{apb}.

 The initial many-body (electron and lattice) quantum state is: 

\begin{eqnarray}
\Psi(\vec r,\vec R,t_0) \!\!= \!\!\sum_{\vec k,\vec q} a(\vec k,\vec q,t_0) \Phi_{\vec k}(\vec r) \Phi_{\vec q}(\vec R),
\label{nqe}
\end{eqnarray}
with $a(\vec k,\vec q,t_0)$  accounting for an arbitrary superposition,  $\Phi_{\vec k}(\vec r) \equiv \langle \vec r| \hat c_{\vec{k}_1}^\dag...\hat c_{\vec{k}_{N_e}}^\dag|0\rangle$ the Slater determinant with $\vec k=\{\vec{k}_1,...,\vec{k}_{N_e}\}$, and $\Phi_{\vec q}(\vec R)$ the atomic part with $\vec q=\{\vec q_1,\vec q_2,...\}$ representing a phonon base. The Slater determinant of electrons can be expanded in minors giving:
\begin{eqnarray}
\label{packet1}
&&\Psi(\vec r_a,\vec z_a,\vec R,t) = \sum_{\vec k,\vec q}  a(\vec k,\vec q,t) \Phi_{\vec q}(\vec R)    \nonumber\\ 
&& \sum_{\vec k_w} \phi_{\vec{k}_w}(\vec r_a) s_{a,w} \langle \vec z_a| \hat c_{\vec{k}_1}^\dag..c_{\vec{k}_{w-1}}^\dag c_{\vec{k}_{w+1}}^\dag..\hat c_{\vec{k}_N}^\dag|{0}\rangle .
\end{eqnarray}
with $s_{a,w}$ the sign of the $(a,w)$ cofactor. Then, the term in \eref{int} acting on \eref{packet1} is (for more details see \aref{apb}): 
\begin{eqnarray}
\label{packet3}
&& \langle \vec r_a,\vec r_b |\mathcal{\hat H}_{ep,\vec u} |\Psi(t) \rangle =\sum_{e,p} g^{\vec {q}_p}_{\vec k_e} \langle \vec r_a,\vec r_b |\hat  c_{\vec{k}_e+\vec{q}_p}^\dag \hat c_{\vec{k}_e} \nonumber\\ && (\hat  b_{\vec{q}_p}+\hat  b_{-\vec{q}_p}^\dag) |\Psi(t)\rangle  = \sum_{e,p} g^{\vec {q}_p}_{\vec k_e}   \sum_{\vec {k},\vec {q}}   a(\vec k,\vec q,t)\Phi'_{\vec q}(\vec R)\nonumber\\
&& \sum_{\vec k_w} \phi_{\vec{k}_w+\vec{q}_p}(\vec r_a) s_{a,w} \langle \vec z_a| \hat  c_{\vec{k}_1}^\dag..c_{\vec{k}_{w-1}}^\dag c_{\vec{k}_{w+1}}^\dag..\hat  c_{\vec{k}_N}^\dag |{0}\rangle.
\end{eqnarray}
We use $\Phi'_{\vec q}(\vec R)$ to account for the effect of the electron-lattice interaction in the atomic part. 

When conditioning Eqs. (\ref{packet1}) and (\ref{packet3}) to $\{\vec z_a^j[t],\vec R^j[t]\}$, the variable $\vec q^j_p$ is also fixed to some particular values in this $j$-th experiment. The exact (deterministic) description of the electron-lattice interaction would require perfect knowledge of all ions dynamics through  $\vec R^j[t]$. However, since ions are considered here as the \emph{environment} of electrons (they are not explicitly simulated), we introduce their effect stochastically in the equation of motion of electrons in \eref{Pseu_Sch}, ensuring that the probabilities of different phonon modes satisfy some well-known precomputed probabilities\cite{JacoboniBook}. We assume that only one (or none) phonon mode $\vec q^j_p$ becomes relevant at each time. Then, the (envelope) CWF before a collision $t<t_{c}$ is:
\begin{eqnarray}
\label{packet2}
&&\psi_a^j(\vec r_a,t) = \sum_{\vec{k}_w}  f(\vec{k}_w,t)\phi_{\vec{k}_w}(\vec r_a).
\end{eqnarray}
Assuming that $g^{\vec {q}_p^j}_{\vec k_a} \approx g^{\vec {q}_p^j}_{\vec k_{0a}}$ with $\vec k_{0a}$ the central wave vector of the $a$ wave packet, the final (envelope) CWF in \eref{packet3} after the collision $t>t_{c2}$ is $\psi_a^j(\vec r_a,t) \equiv \langle \vec r_a,\vec r_b^j[t] |\mathcal{\hat H}_{ep,\vec u} |\Psi(t) \rangle$ which can be written as (see \aref{apb}): 
\begin{eqnarray}
\label{packet4}
&&\psi_a^j(\vec r_a,t)=  g^{\vec {q}_p^j}_{\vec k_{0a}} \sum_{\vec{k}_w} f(\vec{k}_w,t)\phi_{\vec{k}_w+\vec q^j_p}(\vec r_a) .
\end{eqnarray}
where the $\vec r_a$-dependence in \eref{packet2} and \eref{packet4} is given by $\phi_{\vec{k}_a}(\vec r_a)$ and $\phi_{\vec{k}_w+\vec{q}_p}(\vec r_a)$ respectively, and $f(\vec k_w,t)$ includes all other terms evaluated at $\{\vec z_a^j[t],\vec R^j[t]\}$. 

These results have a simple and intuitive explanation. During the collision, the (Bloch state) quasi-momentum eigenstates that build the wave packet change from $|\vec{k}_a\rangle$ to $|\vec{k}_a+\vec q^j_p\rangle$, while its weight $f(\vec{k}_w,t)$ remains constant. 

We notice that these collisions introduce not only stochastic dynamics in the evolution of the CWF, but also time-irreversibility in the whole simulation, since, in general, $g^{\vec {q}_p}_{\vec k_{0a}}< g^{-\vec {q}_p}_{\vec k_{0a}+\vec {q}_p}$, where positive (negative) $\vec{q}_p$ means phonon absorption (emission). 
\begin{figure}[h!!!]
\centering
\includegraphics[scale=0.36]{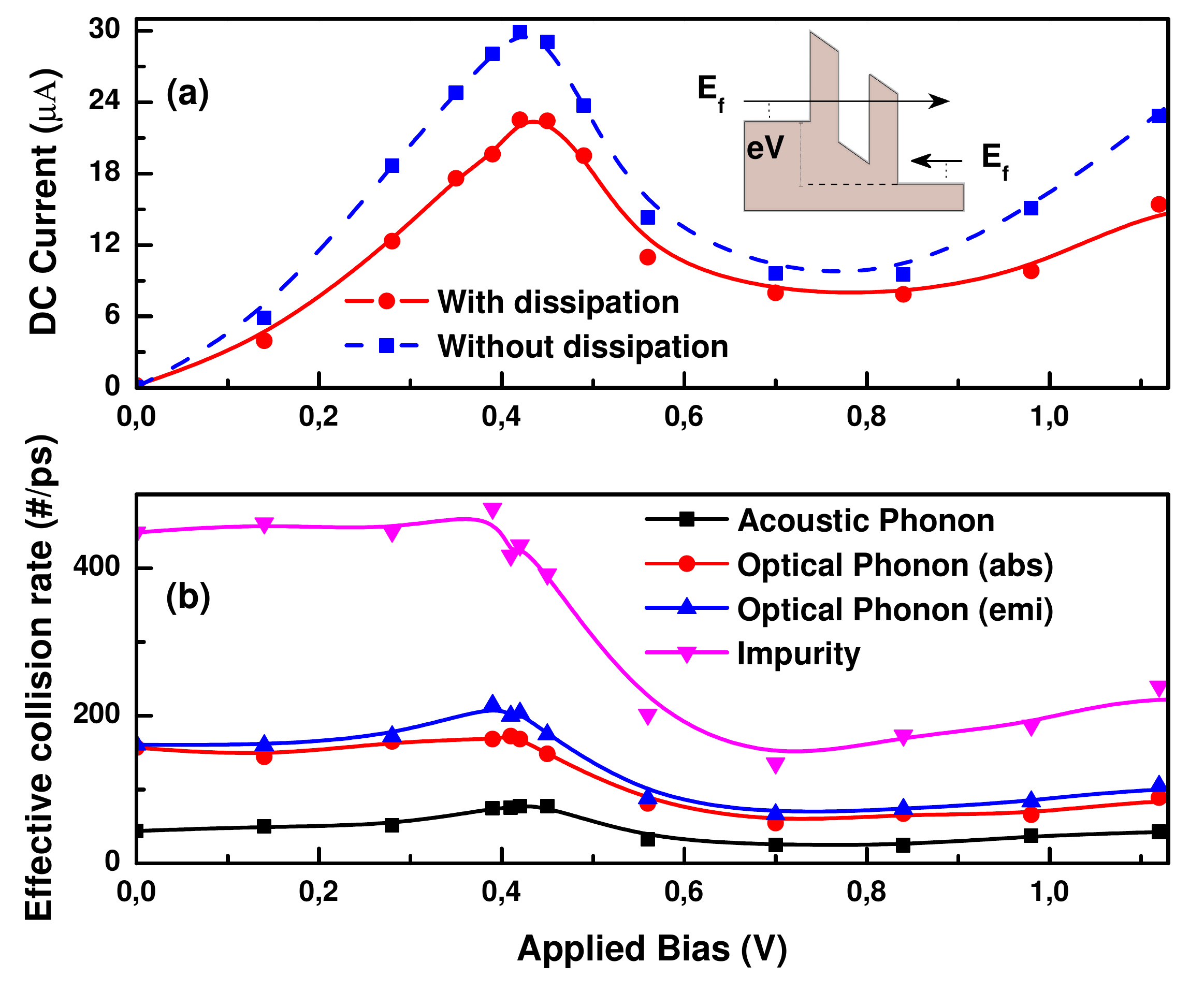}
\caption{(a) Current-voltage characteristic for a RTD with (solid red curve) and without (blue dashed curve) dissipation due to acoustic and optical phonons and impurities. The barrier height and width are $0.5$ eV and $1.6$ nm and the well width is $2.4$ nm. A $n$-type doping with a Fermi level of $E_f=0.15\:eV$  above the conduction band is considered. (b) Effective collision rate as a function of bias. The optical phonons lead to an inelastic change of the electron energy  of $\pm0.036\:eV$. }
\label{fig3}
\end{figure}

\subsection{Dissipative transport in parabolic-band structures}
\label{s3.2}

We apply our approach for the simulation of a typical GaAs/AlGaAs Resonant Tunneling Device (RTD) when  elastic (acoustic phonons and impurities) and inelastic (optical phonons) collisions are considered. In particular it can be shown that the required evolution of the CWF $\psi_a$ interacting with a phonon $\vec q_p=\{q_{px},q_{py},q_{pz}\}$ in a material with parabolic band structure can be obtained from \eref{Pseu_Sch}:  
\begin{eqnarray}
\label{schomany}
 \!\!\!\!i \hbar \frac{\partial \psi_a}{\partial t}\!\! =\!\left [\!\frac{1}{2m^*}\!\!\left( \vec p_a +\vec \lambda_a\Theta_{t_c} \right)^2\!+\!V_a \!\right] \!\psi_a,
\end{eqnarray}
where $\vec p_a=-i\hbar \vec {\nabla}_a$, $m^*=0.067m_e$ is the electron effective mass ($m_e$ is the free electron mass), $\vec \lambda_a=\hbar\vec q_p$ and $\Theta_{t_c} \equiv \Theta(t-t_c)$ is the Heaviside step function.  In \aref{ape} we prove that \eref{schomany} exactly reproduces the transition of $\psi_a$ from \eref{packet2} to \eref{packet4}. Each electron $a=\{1,2,..,N_e\}$ has its own \eref{schomany} to compute $\psi_a$ and $\vec r_a[t]$ by time-integrating its velocity in \eref{velo}. The term $V_a$ provides the Coulomb correlation among all simulated electrons including the appropriate boundaries. The injection model locates the initial CWF outside the simulation box and defines it from typical Gaussian wave packets with a dispersion $\sigma=40$ nm. The properties of the injected electrons are selected according to some well-defined assumptions. For example, the energies of the injected electrons from one contact (assumed in thermodynamical equilibrium) into the open system fulfill a Fermi-Dirac distribution. This randomness in the injection of electrons introduces a source of stochasticity in the description of the properties of the open system.  

We compute the current as the net number of trajectories $\vec r_a[t]$ transmitted from one side to the other, divided by the total simulation time ($5$ ps). Identically the DC current is also computed as the time average of the total (conduction plus displacement) current. Both types of DC computations provide the same value at each bias point, showing the accuracy of the simulation. Technically, the experiment is not repeated, but the numerical simulation takes so long that electrons are entering and leaving the active region many times, providing repeated scenarios. The number and type of collisions are obtained from the Fermi Golden Rule for GaAs materials\cite{JacoboniBook}. We notice that the collision in \eref{Pseu_Sch} does not introduce any artificial decoherence. The expected reduction of the transmission\cite{RTD} seen in \fref{fig3}(a) is because of the randomization of the momentum due to acoustic phonons and to the energy dissipation due to the emission of optical phonons. We see in \fref{fig3}(b) that the number of collisions at resonance is three times larger that out of resonance, showing that the ballisticity of tunneling devices also depends on the electron transit time that varies from one voltage to another, due to different back-actions of our non-Markovian (phonon) environment \cite{ferialdi}.

\subsection{Dissipative transport in linear-band structures}
\label{s3.3}
Next, we present the Bohmian trajectories and CWF evolution of one electron during a collision with a phonon in graphene, with a richer band structure than GaAs. The whole development of the equation of motion in \eref{Pseu_Sch} and the inclusion of the collision needs to be redone for a conditional 2D bispinor $\psi_a\equiv  \left(\psi_{a,1}, \psi_{a,2}\right)^T$ giving:
\begin{eqnarray}
\label{dirac2d2}
i \hbar \frac{\partial \psi_a}{\partial t}\!\! &=&\!\! v_f \!\!\begin{pmatrix}
V_a/v_f & p_a^-+\lambda_a^{-}\Theta_{t_c}\\
(p_a^++\lambda_a^{+}\Theta_{t_c})\chi_{t_c} & V_a\chi_{t_c}/v_f
\end{pmatrix} \psi_a,
\end{eqnarray}
where $v_f\!\!=\!10^6$ m/s is the Fermi velocity. We define $p_a^{\pm}\!=\!-i\hbar\partial_{x_a}\pm\hbar\partial_{y_a}$ and $\lambda^{\pm}_a\!=\!\lambda_{ax}\pm i\lambda_{ay}$ as the change in momentum $\vec \lambda_a=\hbar \vec q_p$ due to the interaction with a phonon with wave vector $\vec q_p=\{q_{px},q_{py}\}$. When the interaction occurs, the term $\chi_{t_c}=e^{i(m\pi+\beta_{\vec {k}_{fa}}-\beta_{\vec {k}_{0a}})\Lambda_{t_c}}$ makes sure that the final state is either in the conduction band (positive energy branch) or in the valence band (negative energy states). If the electron changes from the conduction to the valence band (or vice versa), we use $m=1$ and if there is no change of band $m=0$, with $e^{i\beta_{\vec {k}_{0a}}}=({k_{0ax}+ik_{0ay}})/{|\vec {k}_{0a}}|$, where $\vec {k}_{0a}$ ($\vec {k}_{fa}$) is the central initial (final) wave vector and $e^{i\beta_{\vec {k}_{fa}}}$ having the same definition. $\chi_{t_c}$ is only relevant at $t_c$, i.e. $\Lambda_{t_c}\equiv \Lambda_{t_c}(t)=0$  except $\Lambda_{t_c}(t=t_c)=1$. In \aref{apf} we prove that \eref{dirac2d2} produces the transition of the 2D bispinor $\psi_a$ from \eref{packet2} to \eref{packet4}.

\begin{figure}[ht!!!]
\centering
\includegraphics[scale=0.33]{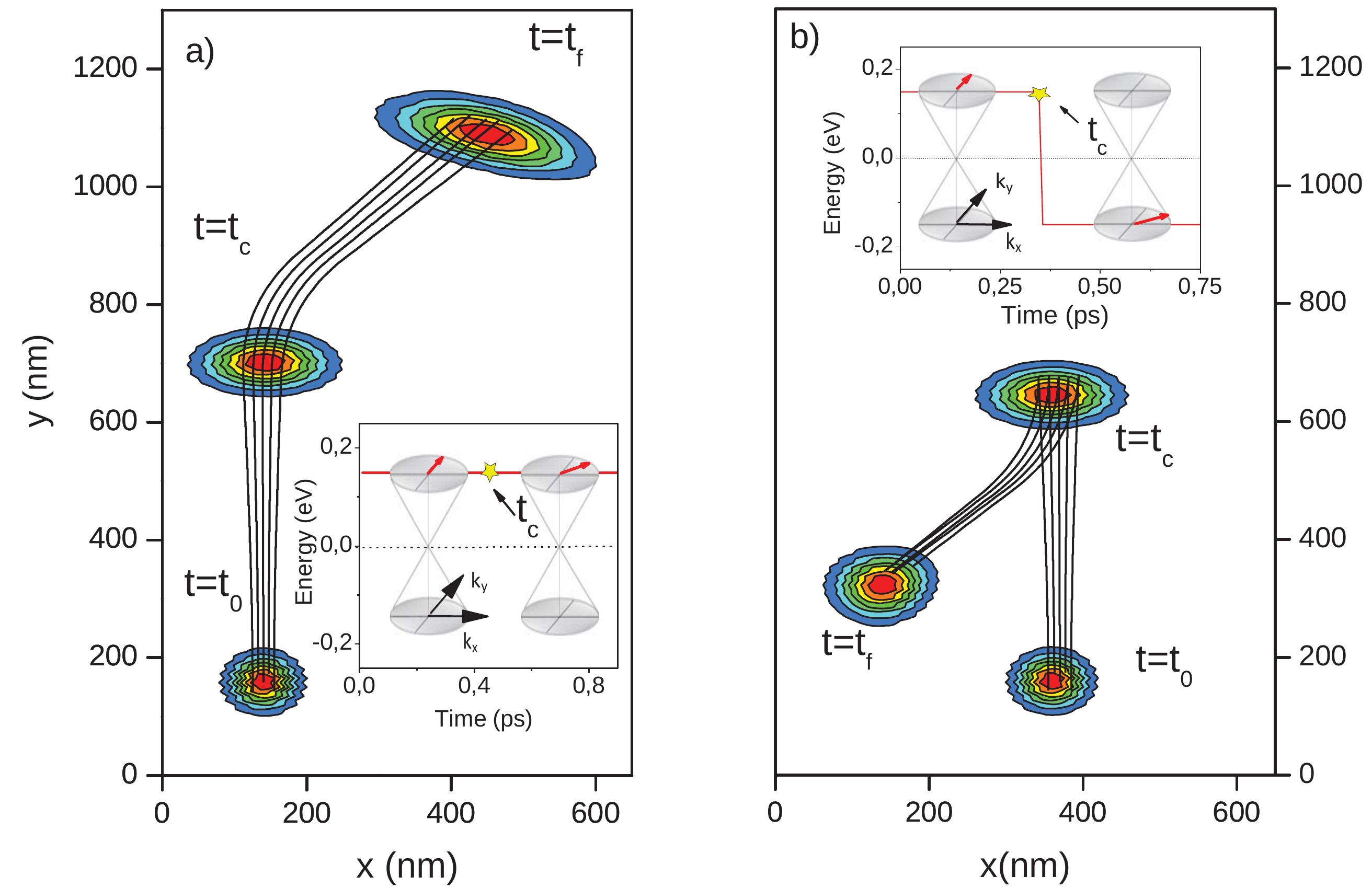}
\caption{a) Time evolution of the modulus squared of the conditional bispinor for an electron initially at $t_0$ in the conduction band, with wave vector $(k_{0ax},k_{0ay})=(0,|\vec{k}_{0a}|)$, suffering an elastic collision at $t_{c}$  with a phonon that provides a final wave vector $(k_{fax},k_{fay})=({|\vec{k}_{0a}|}/{\sqrt{2}},{|\vec{k}_{0a}|}/{\sqrt{2}})$. The associated Bohmian trajectories are also shown. Inset: Electron energy conservation for the elastic collision. b) Same change of wave vector as in a) but with an inelastic collision that produces a final electron in the valence band (where velocity and momentum are opposite).}
\label{wpg}
\end{figure}

\begin{figure}[ht!!!]
\centering
\includegraphics[scale=0.31]{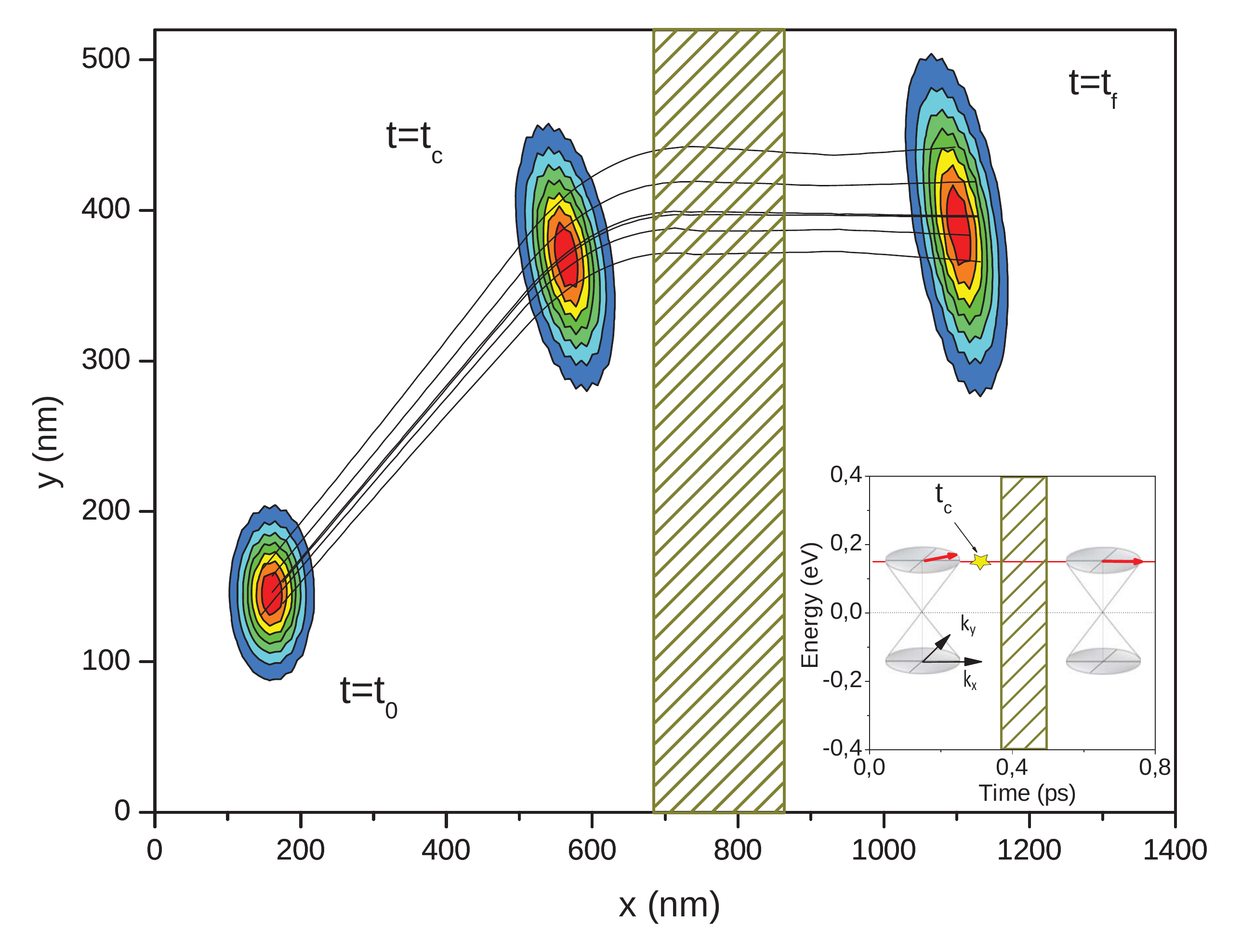}
\caption{Time evolution of the modulus squared of the conditional bispinor for an electron which impinges on a $0.4\:eV$ barrier with a width of $200\:nm$. The initial $t=t_0$ direction is  $\beta_0={\pi}/{6}$ so Klein tunneling should be minimal. At $t=t_{c}$, an elastic collision deviates the electron in a perpendicular direction to the barrier maximizing the Klein tunneling.}
\label{kt}
\end{figure}

We present in \fref{wpg} and \fref{kt} numerical results for the electron-phonon collisions in graphene, whose dynamics near the Dirac points are given by \eref{dirac2d2}. The initial state in both examples is a Gaussian wave packet with dispersion $\sigma=40\:nm$ and wave vector $|\vec{k}_{0a}|=2.27\cdot 10^8\;m^{-1}$, whose initial pseudospin lies in the conduction band.  

\section{Conclusions}
\label{s6}
In this paper we present an approach to analyze quantum dissipation. It is based on Bohmian CWFs that preserves CP and allows a realistic consideration of dissipative sources. Formally, our approach follows the SSE technique\cite{SSE} for non-Markovian scenarios \cite{Gambetta,wiseman,vega}, but allowing a physical interpretation of the output results under a continuous measurement. The open system techniques mentioned in the Introduction are rarely applied to the simulations of electron devices (with exceptions such as the Wigner-Boltzmann approach, which has problems of CP \cite{positivity,Zhen}, or other density matrix approaches that have difficulties in being adapted to spatially well-defined models respecting the different spatial regions (with well-defined boundaries) typical in electron devices \cite{rossi1,rossi2}). Typically, dissipation in quantum electron transport is simulated through a partition of the full Hilbert space into smaller spaces where sets of eigenstates are perfectly determined. Interactions (dissipation) between different spaces are introduced through coupling constants \cite{electdiss}. The solution of such models implicitly involves an improper mixture of states \cite{improper} that, in spite of its computational interest, has no ontological definition within standard quantum mechanics \cite{proper}. The Bohmian CWFs provide a \emph{unproblematic} way to define the wave function of an open system \cite{Gambetta}, and we have shown that it allows a realistic simulation of quantum dissipation in electron devices with linear and parabolic band structures. With the accurate inclusion of quantum dissipation in the evolution of CWFs, the \emph{general} and \emph{accurate} quantum-trajectory approach \cite{BITLLES} presented here is the best candidate to substitute the old Monte Carlo solution of the Boltzmann equation for semi-classical systems \cite{JacoboniBook} in the new nanoelectronics/atomtronics quantum scenarios.

\begin{acknowledgments}
This work has been partially supported by the Fondo Europeo de Desarrollo Regional (FEDER) and the ``Ministerio de Ciencia e Innovaci\'{o}n'' through the Spanish Project TEC2015-67462-C2-1-R, by the Generalitat de Catalunya (2014 SGR-384), and it has also received funding from the EU’s Horizon 2020 research program under grant agreement no: 696656. Z. Zhan acknowledges financial support from the China Scholarship Council (CSC) with No.201306780019.
\end{acknowledgments}


\appendix

\section{Evaluation of the term  $\langle \vec r_a,\vec z_a,\vec R | \mathcal{\hat H}_{ep,\vec u} |\Psi(t) \rangle|_{\vec z_a^j[t],\vec R^j[t]}$ in \eref{condi}:}
\label{apb}
We first evaluate the effect of the $\mathcal{\hat H}_{ep,\vec u}$ on the wave packet  $\Psi(\vec r_a,\vec z_a,\vec R,t)$.  For that purpose, we develop the explicit expression of $\mathcal{\hat H}_{ep,\vec u}$ and then we define an initial many-body wave packet  $\Psi(\vec r_a,\vec z_a,\vec R,t)$.

\subsection*{A1.- Definition of the Electron-Phonon Hamiltonian:}

The term $ H_{ep}$ in \eref{Hto} can be written as $ H_{ep}=H_{ep,\vec R_0}+ H_{ep,\vec u}=\sum_{e,h}{V}_{ep}(\vec r_e-\vec R_h)$. We decompose ${V}_{ep}(\vec r_e-\vec R_h)$ in a Taylor expansion around the equilibrium position  of the $h$ ions $\vec R_{h,0}$ as:
\begin{eqnarray}
\label{int3}
&& \sum_{e,h} {V}_{ep}(\vec r_e-\vec R_h)\approx \sum_{e,h} {V}_{ep}(\vec r_e-\vec R_{h,0})+\\ \nonumber 
&&(\vec R_h\!-\!\vec R_{h,0})\cdot \nabla_h {V}_{ep}(\vec r_e-\vec R_{h})|_{\vec R_{h}=\vec R_{h,0}}=\sum_{e,h}{V}_{ep}(\vec r_e-\vec R_{h,0})\\ \nonumber 
&&+ \vec u_h\cdot \nabla_h {V}_{ep}(\vec r_e-\vec R_{h})|_{\vec R_{h}=\vec R_{h,0}}= H_{ep,\vec R_0}+H_{ep,\vec u}.
\end{eqnarray}
The  term ${V}_{ep}(\vec r_e-\vec R_{h,0})$ will become later relevant for the electronic band structure, while $\vec u_h\cdot \nabla {V}_{ep}(\vec r_e-\vec R_{h})|_{\vec R_{h}=\vec R_{h,0}}$ provides the interaction of the electron $\vec r_e$ with the ion $\vec R_h$ (neglecting second order Taylor terms in the atomic displacements expansion). Instead of dealing with individual displacements $\vec u_h$, we consider the normal coordinate $\vec Q_{\vec {q}_p}$ defined from the Fourier transform:
\begin{eqnarray}
\label{pho1}
\vec u_h =\sum_{\vec q_p} \vec Q_{\vec {q}_p} e^{i\vec {q}_p\vec R_{h,0}},
\end{eqnarray}
where $\vec {q}_p$ is a wave vector in the reciprocal space that labels each of the possible collective solutions of the movement of ions.  Then we perform the Fourier transform of the potential  ${V}_{ep}(\vec r_e-\vec R_{h})$:
\begin{eqnarray}
{V}_{ep}(\vec r_e-\vec R_{h})=\sum_{\vec v} e^{i\vec v( \vec r_e-\vec R_{h})} U_{\vec v},
\label{pot1}
\end{eqnarray}
where $\vec v$ is another wave vector in the reciprocal space and $U_{\vec v}$ is the Fourier coefficients of the potential. Notice that $\sum_h V_{ep}(\vec r_e-\vec R_{h})$ is a periodic potential, while $V_{ep}(\vec r_e-\vec R_{h})$ alone is essentially a Coulomb potential with corrections due to screening. Then, the gradient of the potential in \eref{int3} can be written as: 
\begin{eqnarray}
\nabla_h {V}_{ep}(\vec r_e\!\!-\!\!\vec R_{h})|_{\vec R_{h}=\vec R_{h,0}}\!\!=\!\!\sum_{\vec v} (-i \vec v )e^{i\vec v( \vec r_e-\vec R_{h,0})} U_{\vec v}.
\label{pot2}
\end{eqnarray}
Putting \eref{pot2} and \eref{pho1} altogether for all electrons and ions, we get finally:

\begin{eqnarray}
&& H_{ep,\vec u}= \sum_e H_{ep,\vec u,\vec r_e}=\sum_{e}\sum_{h} \vec u_h\cdot \nabla_h {V}_{ep}(\vec r_e-\vec R_{h})|_{\vec R_{h,0}}\nonumber\\ &&=\sum_e\sum_h \sum_{\vec {q}_p} e^{i \vec {q}_p \vec R_{h,0}} \vec Q_{\vec {q}_p} \sum_{\vec v} (-i \vec v )e^{i\vec v( \vec r_e-\vec R_{h,0})} U_{\vec v}.
\label{aint1}
\end{eqnarray}

Before discussing the interaction through the term $H_{ep,\vec u}$, we define the initial electron-lattice wave packet. 

\subsection*{A2.- Definition of the many body wave packet $\Psi(\vec r,\vec R,t)$:}

The many body wave packet $\Psi(\vec r,\vec R,t)=\langle \vec r_a,\vec z_a,\vec R | \Psi(t) \rangle$ can be written as:
\begin{equation}
\Psi(\vec r,\vec R,t) \!\!= \!\!\sum_{\vec k,\vec q} a(\vec k,\vec q,t) \Phi_{\vec k}(\vec r) \Phi_{\vec q}(\vec R),
\end{equation}
with $a(\vec k,\vec q,t)$ accounting for an arbitrary superposition of the many-body electron base $\Phi_{\vec k}(\vec r)$ and many-body phonon base $\Phi_{\vec q}(\vec R)$. The vector $\vec k=\{\vec k_1,\vec k_2,..,\vec k_{N_e}\}$ represents the many body index of the electronic (Bloch states) base and $\vec q=\{\vec q_1,\vec q_2,...\}$ the index of the ionic base. We define $\Phi_{\vec k}(\vec r) \equiv \sum_{n=1}^{{N_e}!}\prod_{i=1}^{N_e}\phi_{\vec k_i}(\vec r_{p(n)_i})s_n$, with $\vec p(n)=\{p(n)_1,...,p(n)_{N_e}\}$ the $n$-permutation vector, and $s_n$ its sign. We have also used the single particle Bloch eigenstate:
\begin{equation}
\phi_{\vec k_e}(\vec r_e)=\langle \vec r_e | \vec k_e \rangle ={e^{i\vec k_e \vec r_e}}u_{\vec k_e}(\vec r_e),
\label{bloch} 
\end{equation}
where $u_{\vec k_e}(\vec r_e)$ is periodic with respect to lattice translations (which includes the appropriate normalizing constant) and $\vec k_e$ is the electron (quasi) wave vector related to the quasi (or crystal) momentum $\vec p_e=\hbar \vec k_e$.

In the language of the second quantization, the Slater determinant of the electrons can be written as $\Phi_{\vec k}(\vec r) \equiv \langle \vec r| \hat c_{\vec{k}_1}^\dag...\hat c_{\vec{k}_{N_e}}^\dag|0\rangle$. To explicitly write the dependence of $\Phi_{\vec k}(\vec r)$ on $\vec r_e$, we expand the Slater determinant of electrons by minors as  $\langle \vec r_e,\vec z_e| \hat c_{\vec{k}_1}^\dag...\hat c_{\vec{k}_{N_e}}^\dag|0\rangle=\sum_{w=1}^{N_e} \phi_{\vec{k}_w}(\vec r_e) s_{e,w} \langle \vec z_e| \hat c_{\vec{k}_1}^\dag...c_{\vec{k}_{w-1}}^\dag c_{\vec{k}_{w+1}}^\dag...\hat c_{\vec{k}_{N_e}}^\dag|{0}\rangle$, with $s_{e,w}$ the sign of the $(e,w)$ cofactor. Then:
\begin{eqnarray}
\label{apacket1}
&& \Psi(\vec r_e,\vec z_e,\vec R,t) = \sum_{\vec k,\vec q}  a(\vec k,\vec q,t) \Phi_{\vec q}(\vec R)\nonumber\\ && \sum_{\vec k_w} \phi_{\vec{k}_w}(\vec r_e) s_{e,w} \langle \vec z_e| \hat c_{\vec{k}_1}^\dag..c_{\vec{k}_{w-1}}^\dag c_{\vec{k}_{w+1}}^\dag..\hat c_{\vec{k}_N}^\dag|{0}\rangle .
\end{eqnarray}

\subsection*{A3.- Evaluation of $H_{ep,\vec u}(\vec r,\vec R,t) \Psi(\vec r,\vec R,t)$:}

The term  $H_{ep,\vec u}= \sum_e H_{ep,\vec u,\vec r_e}$ in \eref{aint1} is a sum over terms that depend on a unique $\vec r_e$, so that when conditioning $H_{ep,\vec u}(\vec r,\vec R,t) \Psi(\vec r,\vec R,t)$ to $\{\vec z_a^j[t],\vec R^j[t]\}$ all, except one term, do not depend on $\vec r_a$. We have:

\begin{eqnarray}
&&H_{ep,\vec u}(\vec r,\vec R,t)\Psi(\vec r_a,\vec z_a,\vec R,t)\Big|_{\vec z_a^j[t],\vec R^j[t]}\nonumber\\
&&=\left( \sum_e H_{ep,\vec u,\vec r_e}(\vec r_e,\vec R,t)\right)  \Psi(\vec r_a,\vec z_a,\vec R,t)\Big|_{\vec z_a^j[t],\vec R^j[t]}\nonumber\\
&&=\left( \sum_{e\neq a} H_{ep,\vec u,\vec r_e}(\vec r_e^j[t],\vec R^j[t],t)\right)  \Psi(\vec r_a,\vec z_a^j[t],\vec R^j[t],t)\nonumber\\
&&+H_{ep,\vec u,\vec r_a}(\vec r_a,\vec R^j[t],t)\Psi(\vec r_a,\vec z_a^j[t],\vec R^j[t],t).
\end{eqnarray}

The term $\sum_{e\neq a} H_{ep,\vec u,\vec r_e}(\vec r_e^j[t],\vec R^j[t],t)$ is a constant value without dependence on $\vec r_a$. This pure time-dependent term only provides a global phase on the conditional wave function that can be omitted without any effect \cite{PRLxavier}. The only term that we have to compute explicitly is $H_{ep,\vec u,\vec r_a}(\vec r_a,\vec R,t)\Psi(\vec r,\vec R,t) =\langle \vec r, \vec R | \mathcal{\hat H}_{ep,\vec u,\vec r_a} |\Psi(t) \rangle$. Using the identities $\int_{\vec r} d\vec r|\vec r\rangle \langle \vec r|=\mathbb{1}$  and $\int_{\vec R} d\vec R|\vec R\rangle \langle \vec R|=\mathbb{1}$, the fact that $\mathcal{\hat H}_{ep,\vec u,\vec r_a}$  is diagonal in the position representation, and the identity $\sum_{\vec k_a} |\vec k_a \rangle \langle \vec k_a|=\mathbb{1}$, we can write:
\begin{eqnarray}
\label{int22}
&&\langle \vec r, \vec R | \mathcal{\hat H}_{ep,\vec u,\vec r_a} |\Psi(t) \rangle=\langle \vec r, \vec R | \mathcal{\hat H}_{ep,\vec u,\vec r_a} |\vec r, \vec R\rangle \langle \vec r,\vec R |\Psi(t) \rangle=\nonumber \\ 
&&\sum_{\vec k_a} \sum_{\vec k_a''} \langle \vec {r}_a|\vec k_a'' \rangle \langle \vec k_a'', \vec z_a, \vec R | \mathcal{\hat H}_{ep,\vec u,\vec r_a} | \vec k_a,\vec z_a,\vec R \rangle \langle \vec k_a ,\vec z_a, \vec R |\Psi(t) \rangle\nonumber\\
&&=\sum_{\vec k_a} \sum_{\vec k_a''} \mathcal T (\vec k_a'', \mathcal{\hat H}_{ep,\vec u,\vec r_a}, \vec k_a) \langle \vec {r}_a|\vec k_a'' \rangle \langle \vec k_a ,\vec z_a, \vec R |\Psi(t) \rangle,
\end{eqnarray}
where we have defined $ \mathcal T(\vec k_a'', \mathcal{\hat H}_{ep,\vec u,\vec r_a}, \vec k_a)\equiv \langle \vec k_a'', \vec z_a,\vec R |\mathcal{\hat H}_{ep,\vec u,\vec r_a} | \vec k_a,\vec z_a,\vec R \rangle$ as the electron-phonon Hamiltonian in the momentum (Bloch state) representation. This term can be rewritten as:
\begin{eqnarray}
\label{int23}
&& \mathcal T(\vec k_a'', \mathcal{\hat H}_{ep,\vec u,\vec r_a}, \vec k_a)= \\ 
&& \int_{\vec r_a} \!\!\!d \vec r_a \langle \vec k_a'' | \vec r_a\rangle \langle \vec r_a, \vec z_a, \vec R |\mathcal{\hat H}_{ep,\vec u,\vec r_a}| \vec r_a,\vec z_a,\vec R \rangle  \langle \vec r_a | \vec k_a \rangle,\nonumber
\end{eqnarray} 
and using the final expression of the electron-phonon Hamiltonian in the position representation, \eref{aint1}, we get:
\begin{eqnarray}
\label{int33}
&&\mathcal T(\vec k_a'', \mathcal{\hat H}_{ep,\vec u,\vec r_a}, \vec k_a) =\int_{\vec r_a} d \vec r_a {e^{-i\vec k_a'' \vec r_a}}\nonumber\\
&&u_{\vec k_a''}(\vec r_a) {e^{i\vec k_a \vec r_a}}u_{\vec k_a}(\vec r_a) \langle \vec r_a, {\vec z_a}, \vec R |\mathcal{\hat H}_{ep,\vec u,\vec r_a} | \vec r_a, \vec z_a,\vec R \rangle = \nonumber\\
&&\int_{\vec r_a} d \vec r_a {e^{-i\vec k_a'' \vec r_a}}u_{\vec k_a''}(\vec r_a) {e^{i\vec k_a \vec r_a}}u_{\vec k_a}(\vec r_a)\sum_h \sum_{\vec q_p} e^{i \vec {q}_p \vec R_{h,0}} \vec Q_{\vec {q}_p} \nonumber\\
&&\sum_{\vec v} (-i \vec v )e^{i\vec v( \vec r_a-\vec R_{h,0})} U_{\vec v}.
\end{eqnarray}
We take away from the integral those elements that do not depend on $\vec r_a$:
\begin{eqnarray}
&&\mathcal T(\vec k_a'', \mathcal{\hat H}_{ep,\vec u,\vec r_a}, \vec k_a)=\sum_h \sum_{\vec {q}_p} e^{i \vec {q}_p \vec R_{h,0}} \vec Q_{\vec {q}_p} \\
&&\sum_{\vec v} (-i \vec v )e^{-i\vec  v\vec R_{h,0}} U_{\vec v}\int_{\vec r_a} d \vec r_a {e^{-i\vec k_a'' \vec r_a}}u_{\vec k_a''}(\vec r_a) {e^{i\vec k_a \vec r_a}}u_{\vec k_a}(\vec r_a) e^{i\vec v \vec r_a}.\nonumber
\label{int4}
\end{eqnarray}
Due to the periodicity of $u_{\vec k_a}(\vec r_a)$ we can use the change of variable $\vec r_a=\vec {r'}_a+\vec R_{m,0}$ where $\vec {r'}_a$ integrates only inside the first Brillouin zone. We get: 
\begin{eqnarray}
&&\mathcal T(\vec k_a'', \mathcal{\hat H}_{ep,\vec u,\vec r_a}, \vec k_a)=\nonumber\\
&& \sum_h  \sum_{\vec {q}_p}  \vec Q_{\vec {q}_p} \sum_{\vec v} (-i \vec v )e^{i \vec R_{h,0} (\vec {q}_p -\vec v)} U_{\vec v}\left(\sum_m e^{i\vec R_{m,0} (-\vec k_a''+\vec v+\vec k_a)}\right)\nonumber\\
&&\int_{\vec {r'}_a} d \vec {r'}_a {e^{-i\vec k_a'' \vec {r'}_a}}u_{\vec k_a''}(\vec {r'}_a) {e^{i\vec k_a \vec {r'}_a}}u_{\vec k_a}(\vec {r'}_a) e^{i\vec v \vec {r'}_a}.
\label{int5}
\end{eqnarray}

The sum over $\vec R_{h,0}$ in $\sum_h e^{i \vec R_{h,0} (\vec {q}_p -\vec v)}$ imposes the condition $\vec G=\vec {q}_p -\vec v$ and the sum over $\vec R_{m,0}$ in $\sum_m e^{i\vec R_{m,0} (-\vec k_a''+\vec v+\vec k_a)}$ imposes that $\vec{G'}=-\vec k_a''+\vec {v}+\vec k_a$, with $\vec G$ and $\vec G'$ two vectors of the reciprocal lattice. For simplicity, although Umklapp scattering can also be considered, we assume that all momentum vectors can be considered in the first Brillouin zone, $\vec G=0$ and $\vec G'=0$, so that $\vec k_a''=\vec {q}_p+\vec k_a$. Therefore: 
\begin{eqnarray}
\mathcal T(\vec k_a'', \mathcal{\hat H}_{ep,\vec u,\vec r_a}, \vec k_a)=\sum_{\vec {q}_p}\delta (\vec k_a''-\vec {q}_p -\vec k_a) g^{\vec {q}_p}_{\vec k_a}.
\label{int6}
\end{eqnarray}
All other terms in \eref{int5} are included into the coupling constant $g^{\vec {q}_p}_{\vec k_a}$ defined as:
\begin{eqnarray}
&& g^{\vec {q}_p}_{\vec k_a}=-i \vec Q_{\vec {q}_p} \; \vec {q}_p \; U_{\vec {q}_p} \int_{\vec {r'}_a} d \vec {r'}_a {e^{-i(\vec k_a+\vec q_p) \vec {r'}_a}}\nonumber\\
&&u_{\vec k_a+\vec q_p}(\vec {r'}_a) {e^{i\vec k_a \vec {r'}_a}}u_{\vec k_a}(\vec {r'}_a) e^{i\vec v \vec {r'}_a}.
\label{g_par}
\end{eqnarray}
We emphasize that we did not include any dependence on the $n$ band we are dealing with, since usually phonon energies are smaller than band gaps and then phonons cannot make band transitions. However, for materials with small band gaps these multi band transitions can be included straightforwardly. In fact, when dealing with graphene bispinors, we will include the dependence of the coupling constant on the energy branches. We introduce \eref{int6} into \eref{int22} and we conclude:
\begin{eqnarray}
\label{int7}
&& H_{ep,\vec u,\vec r_a}(\vec r_a,\vec z_a,\vec R,t) \Psi(\vec r_a,\vec z_a,\vec R,t)=\nonumber\\
&&\sum_{\vec k_a}  \sum_{\vec k_a''}   \langle \vec r_a|\vec k_a''\rangle \mathcal T(\vec k_a'', \mathcal{\hat H}_{ep,\vec u,\vec r_a}, \vec k_a)\langle \vec k_a,\vec z_a^j[t],\vec R^j[t]  |\Psi(t) \rangle \nonumber\\
&&=\sum_{\vec {q}_p} \sum_{\vec k_a}  g^{\vec {q}_p}_{\vec k_a} \langle \vec r_a|\vec k_a +\vec {q}_p \rangle \langle \vec k_a,\vec z_a,\vec R |\Psi(t) \rangle.
\end{eqnarray}
 
\subsection*{A4.- Conditional (envelope) wave packet before the collision:}

The conditional wave packet before the collision can be obtained from \eref{apacket1} by fixing $\vec z_a=\vec z_a^j[t]$ and $\vec R=\vec R^j[t]$ where these positions correspond to one $j$ experiment. Then:
\begin{eqnarray}
\label{packet2a}
&&\Psi(\vec r_a,\vec z_e^j[t],\vec R^j[t],t) = \sum_{\vec k,\vec q}  a(\vec k,\vec q,t) \Phi_{\vec q}(\vec R^j[t]) \nonumber\\
&&\sum_{\vec k_w} \phi_{\vec{k}_w}(\vec r_a) s_{a,w} \langle \vec z_a^j[t]| \hat c_{\vec{k}_1}^\dag..c_{\vec{k}_{w-1}}^\dag c_{\vec{k}_{w+1}}^\dag..\hat c_{\vec{k}_N}^\dag|{0}\rangle .
\end{eqnarray}
The dependence on $\vec r_a$ of the conditional wave packet in \eref{packet2a} appears because of the Bloch state $\phi_{\vec{k}_w}(\vec r_a) \equiv\langle \vec r_a|\vec k_w \rangle$. Therefore, it  can be compactly rewritten as:
\begin{eqnarray}
\label{intbis}
&& \psi_a(\vec r_a,t)\equiv \Psi(\vec r_a,\vec z_a^j[t], \vec R^j[t],t) \\
&&=\sum_{\vec k_w}  f_a(\vec{k}_w,\vec z_a^j[t], \vec R^j[t],t)\phi_{\vec{k}_w}(\vec r_a)=\sum_{\vec k_w}f_a(\vec{k}_w,t)  \phi_{\vec{k}_w}(\vec r_a),\nonumber
\end{eqnarray}
where $f_a(\vec k_w,t) \equiv f_a(\vec{k}_w,\vec z_a^j[t], \vec R^j[t],t) \equiv \langle \vec k_w ,\vec z_a^j[t],\vec R^j[t] |\Psi(t) \rangle$, appearing in \eref{packet2} and \eref{packet4} in the text, is defined here as:

\begin{eqnarray}
\label{fw}
&& f_a(\vec{k}_w,t) =\sum_{\vec q} \sum_{\vec k,\vec k_e\neq \vec k_{w}}  a(\vec k,\vec q,t) \Phi_{\vec q}(\vec R^j[t]) s_{a,w} \nonumber \\ &&\langle \vec z_a^j[t]| \hat c_{\vec{k}_1}^\dag .. c_{\vec{k}_{w-1}}^\dag c_{\vec{k}_{w+1}}^\dag .. \hat c_{\vec{k}_N}^\dag|{0}\rangle .
\end{eqnarray}

Under the standard envelope approximation in which the wave packet is centered around $\vec k_a \approx \vec k_{0a}$, we can rewrite the Bloch states as $\langle \vec r_a | \vec k_a \rangle = \phi_{\vec k_a}(\vec r_a) \approx {e^{i\vec k_a \vec r_a}}u_{\vec k_{0a}}(\vec r_a)$ and rewrite \eref{intbis} as:
\begin{eqnarray} 
&&\psi_a(\vec r_a,t) =u_{\vec k_{0a}}(\vec r_a) \sum_{\vec k_w}  {e^{i\vec k_w\vec r_a}} \langle \vec k_w ,\vec z_a^j[t],\vec R^j[t] |\Psi(t) \rangle \nonumber\\
&&=u_{\vec k_{0a}}(\vec r_a) \sum_{\vec k_w}  {e^{i\vec k_w\vec r_a}} f_a(\vec k_w ,t).
\label{before}
\end{eqnarray}
We notice that $f(\vec{k}_a,t)$ includes now an (irrelevant) normalization constant. Finally, we notice that we will use the same symbol $\psi_a(\vec r_a,t)$ to refer to the conditional wave packet and to the envelope conditional wave function defined, by ignoring the atomic periodicity $u_{\vec k_{0a}}(\vec r_a)$, as:
\begin{eqnarray} 
&&\psi_a(\vec r_a,t) = \sum_{\vec k_w}  {e^{i\vec k_w\vec r_a}} \langle \vec k_w ,\vec z_a^j[t],\vec R^j[t] |\Psi(t) \rangle \nonumber\\
&&= \sum_{\vec k_w}  {e^{i\vec k_w\vec r_a}} f_a(\vec k_w ,t),
\label{enbefore}
\end{eqnarray}
The ensemble momentum of the initial envelope wave packet $\psi_a(\vec r_a,t)$ in \eref{enbefore}, at $t=t_{c1}$ before the collision, can be written as:
\begin{eqnarray}
\label{intmm}
\langle \vec p_a \rangle_{t_{c1}}=\sum_{\vec k_w} \hbar \vec k_w |f_a(\vec k_w,t)|^2.
\end{eqnarray}

\subsection*{A5.- Conditional (envelope) wave packet after the collision:}
Conditioning the many-body wave function $\Psi(\vec r_a,\vec z_e,\vec R,t)$ to the particular values of $\vec z_a^j[t]$ and $\vec R^j[t]$ belonging to the $j$ experiment means that we are considering only one event of the many available in the wave function. In particular, from all phonon modes present in \eref{int7}, we consider that, in a particular $j$ experiment, only one $\vec q_p^j[t]$ (or none) is relevant at each time $t$ (if more than one phonon mode is relevant simultaneously then we can assume two single-phonon collisions simultaneously, each one with only one type of phonon mode). In addition, we consider that the involved wave packets are narrow enough in momentum space so that $g^{\vec {q}_p,j}_{\vec k_a}[t] \approx g^{\vec {q}_p,j}_{\vec k_{0a}}[t]$, with $\vec k_{0a}$ the central wave vector of the $a$ wave packet. Then, \eref{int7} conditioned to the value of $\vec z_a^j[t]$ and $\vec R^j[t]$ can be written as:
\begin{eqnarray}
\label{int8}
&& H_{ep,\vec u,\vec r_a}(\vec r_a,\vec z_a,\vec R,t) \Psi(\vec r_a,\vec z_a,\vec R,t)\big|_{\vec z_a^j[t]\vec R^j[t]}\\
&&=g^{\vec {q}_p^j}_{\vec k_{0a}} [t] \sum_{\vec k_w} \langle \vec r_a|\vec k_w +\vec {q}_p \rangle \langle \vec k_w,\vec z_a^j[t],\vec R^j[t] |\Psi(t) \rangle.\nonumber
\end{eqnarray}
The coupling constant $g^{\vec {q}_p^j}_{\vec k_{0a}} [t]$ in the $j$ experiment will imply an interaction of the $\vec r_a$ electron with the phonon mode $\vec q_p$ during a collision time interval, starting at $t_{c1}$ and ending at $t_{c2}$. In a later time $g^{\vec {q}_p^j}_{\vec k_{0a}} [t]$ will indicate a collision with another phonon mode. The exact (deterministic) description of $g^{\vec {q}_p^j}_{\vec k_{0a}} [t]$ would require perfect knowledge of the dynamics of $\vec R^j[t]$. Since we do not explicitly simulate the dynamics of the ions (which are understood as the environment of the electrons), we can only introduce their effects in a stochastic way ensuring that the probabilities of different phonon modes given by $g^{\vec {q}_p^j}_{\vec k_{0a}} [t]$ satisfy some precomputed values.  This is the origin of the stochasticity in \eref{Pseu_Sch} due to the environment. 

In one particular $j$ experiment, during one collision, the term $g^{\vec {q}_q}_{\vec k_{0a}}[t]$ becomes irrelevant (the Bohmian velocity does only depends on the dependence of the phase on $\vec r_a$, not on the norm) and the final wave packet in \eref{int8},  at time $t>t_{c2}$ after the collision, can be written as:
\begin{eqnarray}
\label{finbis}
&& \psi_a(\vec r_a,t)\equiv \Psi(\vec r_a,\vec z_a^j[t], \vec R^j[t],t)\nonumber  \\ \nonumber &&=\sum_{\vec k_w}  f_a(\vec{k}_w,\vec z_a^j[t], \vec R^j[t],t)\phi_{\vec{k}_w+\vec q_p}(\vec r_a)\\  &&=\sum_{\vec k_w} f_a(\vec{k}_w,t)\phi_{\vec{k}_w+\vec q_p}(\vec r_a),
\end{eqnarray}
where $f_a(\vec{k}_w,t)=\langle \vec k_w ,\vec z^j_a[t] ,\vec R^j[t] |\Psi(t) \rangle$ remains equal to the value  in \eref{fw} before the collision. 

After the collision at $t=t_{c2}$,  the (pseudo) momentum base changes from $|\vec{k}_e\rangle$ to $|\vec{k}_e+\vec q^j_p\rangle$, so that the final ensemble momentum of the envelope conditional wave packet in \eref{finbis} is:
\begin{eqnarray}
\label{finmm}
&&\langle \vec p_a \rangle_{t_{c2}}=\sum_{\vec k_w} \hbar( \vec k_w+\vec q^j_p) |f_a(\vec k_w,t)|^2\nonumber \\  &&=\langle \vec p_a \rangle_{t_{c1}} +\hbar \vec q^j_p.
\end{eqnarray}

Let us emphasize that \eref{intmm} and \eref{finmm} provide the expected role of the electron-phonon interaction: Such collision generates a change of momentum $\hbar \vec q^j_p$ in the conditional wave packet during a time interval $t_{c2}-t_{c1}$. We notice that we are considering a collision with a finite duration. As it will be later explained, for simplicity in practical applications, we have considered instantaneous collisions in the text.

\section{Evaluation of the term  $\langle \vec r_a,\vec z_a,\vec R | \mathcal{\hat H}_{c} |\Psi(t) \rangle|_{\vec z_a^j[t],\vec R^j[t]}$ in \eref{condi2}:}
\label{apc}

The term $\langle \vec r_a,\vec z_a,\vec R | \mathcal{\hat H}_c|\Psi(t) \rangle|_{\vec z_a^j[t],\vec R^j[t]}=\langle \vec r_a,\vec z_a,\vec R |K_e(\vec r)+ K_h(\vec R)+ V_{ee}(\vec r)+ V_{hh}(\vec R)+ H_{ep,\vec R_0}(\vec r,\vec R_0)|\Psi(t) \rangle|_{\vec z_a^j[t],\vec R^j[t]}$ can be evaluated as follows. First, we divide $V_{ee}(\vec r)= V_{ee, \vec r_a}(\vec r_a)+ V_{ee,\vec z_a}(\vec z_a)$ as the terms with an explicit dependence on $\vec r_a$, plus the terms without it. Similarly, $ H_{ep,\vec R_0}(\vec r,\vec R_0)= H_{ep,\vec R_0,\vec r_a}(\vec r_a,\vec R_0)+ H_{ep,\vec R_0, \vec z_a}(\vec z_a,\vec R_0)$.

\subsection*{B1.- Evaluation of $\langle \vec r_a,\vec z_a,\vec R | \mathcal{\hat V}_{hh}+\mathcal{\hat V}_{ee,\vec z_a} +{\hat H}_{ep,\vec R_0, \vec z_a} |\Psi(t) \rangle|_{\vec z_a^j[t],\vec R^j[t]}$:}

We have:
\begin{eqnarray}
&&\langle \vec r_a,\vec z_a,\vec R | \mathcal{\hat V}_{hh}+\mathcal{\hat V}_{ee,\vec z_a} +{\hat H}_{ep,\vec R_0, \vec z_a} |\Psi(t) \rangle|_{\vec z_a^j[t],\vec R^j[t]}=\nonumber \\  &&\left( V_{hh}(\vec R^j[t]) +{ V}_{ee,\vec z_a}(\vec z_a^j[t]) +{ H}_{ep,\vec R_0, \vec z_a}(\vec z_a^j[t],\vec R_0)\right) \nonumber \\  &&\Psi(\vec r_a,\vec z_a^j[t],\vec R^j[t],t),
\end{eqnarray}
where ${ V}_{hh}(\vec R^j[t])+{ V}_{ee,\vec z_a}(\vec z_a^j[t])+ H_{ep,\vec R_0, \vec z_a}(\vec z_a^j[t],\vec R_0)$ are pure time-dependent terms, without $\vec r_a$ dependence and then it only contributes to an arbitrary pure time-dependent angle that can be directly ignored, see Ref. \onlinecite{PRLxavier}. 

\subsection*{B2.- Evaluation of $\langle \vec r_a,\vec z_a,\vec R | \mathcal{\hat V}_{ee, \vec r_a} |\Psi(t) \rangle|_{\vec z_a^j[t],\vec R^j[t]}$:}

Similarly, we write: 
\begin{eqnarray}
&& \langle \vec r_a,\vec z_a,\vec R | \mathcal{\hat V}_{ee, \vec r_a}\Psi(t) \rangle|_{\vec z_a^j[t],\vec R^j[t]} \nonumber \\  &&= V_{ee, \vec r_a}(\vec r_a,\vec z_a^j[t]) |\Psi(\vec r_a,\vec z_a^j[t],\vec R^j[t],t)\nonumber \\  &&= u_{\vec k_{0a}}(\vec r_a) V_{ee, \vec r_a}(\vec r_a,\vec z_a^j[t]) \psi_a(\vec r_a,t),
\label{t01}
\end{eqnarray}
where $V_{ee,\vec r_a}(\vec r_a,\vec z_a^j[t])$ can be easily known once the set of $\vec r^j[t]$ trajectories are known. Later we will use $V_a \equiv V_{ee,\vec r_a}(\vec r_a,\vec z_a^j[t])$

\subsection*{B3.- Evaluation of $\langle \vec r_a,\vec z_a,\vec R | \mathcal{\hat K}_{e,\vec z_a}+ \mathcal{\hat K}_{h} |\Psi(t) \rangle|_{\vec z_a^j[t],\vec R^j[t]}$:}

The kinetic energy of ions $\mathcal{\hat K}_{h}$ and the kinetic energy of the rest of electrons, different from $\vec r_a$, defined as $\mathcal{\hat K}_{e,\vec z_a}$ with $\mathcal{\hat K}_{e}=\mathcal{\hat K}_{e,\vec r_a}+\mathcal{\hat K}_{e,\vec z_a}$, can be written as:
\begin{eqnarray}
&&\langle \vec r_a,\vec z_a,\vec R | \mathcal{\hat K}_{e,\vec z_a}+ \mathcal{\hat K}_{h} |\Psi(t) \rangle|_{\vec z_a^j[t],\vec R^j[t]}\nonumber \\  &&=\sum_{e=1,e \neq a}^{N_e}{K}_{e,\vec r_a}\Psi(\vec r_a, \vec z_a, \vec R,t)  \big|_{\vec z_a^j[t],\vec R^j[t]} \nonumber \\  &&+\sum_{h=1}^{N_h}\frac{-\hbar^2 }{2m_h}\vec \nabla_h^2 \Psi(\vec r_a, \vec z_a, \vec R,t)  \big|_{\vec z_a^j[t],\vec R^j[t]}\nonumber\\
&&=A(\vec r_a,\vec z_a^j[t],\vec R^j[t],t)\Psi(\vec r_a,\vec z_a^j[t],\vec R^j[t],t)\nonumber \\  &&=u_{\vec k_{0a}}(\vec r_a) A_a(\vec r_a,\vec z_a^j[t],\vec R^j[t],t) \psi_a(\vec r_a,t).
\label{t02}
\end{eqnarray}
where ${K}_{e,\vec r_a}$ is the kinetic energy of each $\vec z_a$ electron and where we have introduced the real potential $A_a$ as:
\begin{eqnarray}
&& A_a \equiv A_a(\vec r_a,\vec z_a^j[t],\vec R^j[t],t)=\nonumber \\  &&
\frac{\sum_{e=1,e \neq a}^{N_e}{K}_{e,\vec r_a} \Psi(\vec r_a, \vec z_a, \vec R,t)  \big|_{\vec z_a^j[t],\vec R^j[t]}}{\Psi(\vec r_a,\vec z_a^j[t],\vec R^j[t],t)}\nonumber \\  &&
+\frac{\sum_{h=1}^{N_h}\frac{-\hbar^2 }{2m_h}\vec \nabla_h^2 \Psi(\vec r_a, \vec z_a, \vec R,t)  \big|_{\vec z_a^j[t],\vec R^j[t]}}{\Psi(\vec r_a,\vec z_a^j[t],\vec R^j[t],t)}.
\label{A}
\end{eqnarray}
This constant $A_a$ includes other correlations (different from the electron-lattice correlations that we treat exactly apart from the stochastic approximation for ions dynamics) and will be approximated later according to Ref.\onlinecite{PRLxavier}.

\subsection*{B4.- Evaluation of $\langle \vec r_a,\vec z_a,\vec R | \mathcal{\hat K}_{e,\vec r_a}+ \mathcal{\hat H}_{ep,\vec R_0,\vec r_a}|\Psi(t) \rangle|_{\vec z_a^j[t],\vec R^j[t]}$:}

The last terms that have to be evaluated from $\mathcal{\hat H}_{c}$ in \eref{condi} are $\mathcal{\hat K}_{e,\vec r_a}+ \mathcal{\hat H}_{ep,\vec R_0,\vec r_a}$. They determine the electronic band structure:  
\begin{eqnarray}
&&\langle \vec r_a,\vec z_a,\vec R | \mathcal{\hat K}_{e,\vec r_a}+ \mathcal{\hat H}_{ep,\vec R_0,\vec r_a} |\Psi(t) \rangle|_{\vec z_a^j[t],\vec R^j[t]} \\  &&=\left({K}_{e,\vec r_a} +\sum_h {V}_{ep}(\vec r_a-\vec R_{h,0})\right)\Psi(\vec r_a, \vec z_a^j[t], \vec R^j[t],t),\nonumber
\label{t030}
\end{eqnarray}
where ${K}_{e,\vec r_a}$ corresponds to the kinetic energy of the conditioned $a$ electron and ${H}_{ep,\vec R_0,\vec r_a}=\sum_h {V}_{ep}(\vec r_a-\vec R_{h,0})$ is the periodic potential seen by this $a$ electron. From here, and after a tight binding and the approximation for low energy excitations (small $\vec k$), depending on the system we will end up with a band structure $E(\vec p)$ either with linear or parabolic shape. Therefore:

\begin{eqnarray}
&&\langle \vec r_a,\vec z_a,\vec R | \mathcal{\hat K}_{e,\vec r_a}+ \mathcal{\hat H}_{ep,\vec R_0,\vec r_a} |\Psi(t) \rangle|_{\vec z_a^j[t],\vec R^j[t]}\nonumber \\  &&\approx u_{\vec k_{0a}}(\vec r_a) E(\vec p) \psi_a(\vec r_a,t),
\label{t03}
\end{eqnarray}
where $\psi_a(\vec r_a,t)$ is the (conditional) envelope wave packet already defined in \eref{before}. 

\section{Sch\"odinger (parabolic band) equation}
\label{ape}

In the parabolic case, $E(\vec p_a)$ appearing in \eref{t03} is $E(\vec p_a)=\frac{|\vec p_a|^2}{2m^*}$ with $m^*$ an isotropic effective mass. After the collision, $t=t_{c2}$, regarding \eref{finbis} the  state $|\vec{k}_e\rangle$ changes to $|\vec{k}_e+\vec q^j_p\rangle$. Under the mentioned envelope approximation, the Bloch states are $\langle \vec r_a | \vec k_a \rangle = \phi_{\vec k_a}(\vec r_a) \approx {e^{i\vec k_a \vec r_a}}u_{\vec k_{0a}}(\vec r_a)$. Then, the conditional wave packet at $t_{c2}$ in \eref{finbis} can be related to the initial wave packet at $t_{c1}$ given by \eref{enbefore} as:
\begin{eqnarray} 
&& \psi_a(\vec r_a,t_{c2}) = \sum_{\vec k_w}  {e^{i(\vec k_w+\vec q_p)\vec r_a}} \langle \vec k_w ,\vec z_a^j[t_{c2}],\vec R^j[t_{c2}] |\Psi(t_{c2}) \rangle \nonumber\\ 
 &&=\sum_{\vec k_w}  {e^{i(\vec k_w+\vec q_p)\vec r_a}} f_a(\vec k_w ,t_{c2})={e^{i\vec q_p\vec r_a}} \psi_a(\vec r_a,t_{c1}).
\label{after}
\end{eqnarray}
Therefore, since Bloch states are energy eigenstates, the ensemble energy before the collision $\langle E(\vec k_a) \rangle_{t_{c1}}$ changes to the value $\langle E(\vec k_a+\vec q_p) \rangle_{t_{c2}}$ after the collision.\\

Putting together \eref{t01}, \eref{t02}, \eref{t03} and \eref{t04} into the original \eref{condi}, and removing $u_{\vec k_{0a}}(\vec r_a) $, we get:
\begin{eqnarray}
\label{schomany1}
 i \hbar \frac{\partial \psi_a(\vec r_a,t)}{\partial t}\!\!=\!\! \left [\frac{1}{2m^*}\!\!\left( \vec p_a \right)^2 \!\!+\!V_a\!\!+\!A_a\!+\!iB_a\!\!\right]\!\! \psi_a(\vec r_a,t),
\end{eqnarray}
where the terms $A_a$ and $B_a$ in \eref{A} and \eref{B} are approximated by a zero order Taylor expansion (i.e. no dependence on $\vec r_a$) so that they can be neglected when computing Bohmian velocities. See in Ref. \onlinecite{PRLxavier} the discussion of such approximation. Therefore, the time evolution operator (propagator) from the initial time $t_0$ until a time before the collision $t<t_{c1}$ is just  $\hat {\mathcal U}_a(t,t_0)=e^{-\frac{i}{\hbar} \int_{t_0}^{t} \hat {\mathcal H}_{ca}(t')dt'}$, with $ H_{ca}=\frac{1}{2m^*}\left( \vec p_a \right)^2 +V_a$, being $H_{c}$ conditioned at $\vec z_a^j[t]$.

The time-evolution of the wave packet due to the collision with the phonon has to reproduce the condition given by \eref{after}. The time evolution operator (propagator) from $t=t_0$ until a time $t>t_{c2}$ after the collision is then:
\begin{eqnarray}
&&\hat {\mathcal U}_a(t,t_0)=e^{-\frac{i}{\hbar} \int_{t_{c2}}^{t} \hat {\mathcal H}_{ca}(t')dt'} e^{-\frac{i}{\hbar} \int_{t_{c1}}^{t_{c2}} \hat {\mathcal H}_{epa}(t')dt'}\\ 
 && e^{-\frac{i}{\hbar} \int_{t_0}^{t_{c1}} \hat {\mathcal H}_{ca}(t')dt'} = e^{-\frac{i}{\hbar} \int_{t_{c2}}^{t} \hat {\mathcal H}_{ca}(t')dt'} e^{i\frac{\vec{\lambda}_a \vec{r}_a }{\hbar}} e^{-\frac{i}{\hbar} \int_{t_0}^{t_{c1}} \hat {\mathcal H}_{ca}(t')dt'},\nonumber
\end{eqnarray}
where $\hat {\mathcal H}_{epa}=-\hbar \vec{\lambda}_a \vec{r}_a \delta(t-t_c)$ is the previously mentioned total electron-lattice interaction $\hat {\mathcal H}_{ep}$ conditioned at $\vec z_a^j[t]$.\\

For a small time interval, $\Delta t$, we have $\hat {\mathcal U}_a(t+\Delta t,t) =(1-\frac{i}{\hbar} \Delta t \hat {\mathcal H}_{ca})$. Then, it can be proven that $(1-\frac{i}{\hbar} \Delta t  H_{ca}) e^{i\frac{\vec{\lambda}_a \vec{r}_a }{\hbar}}\psi_a(\vec r_a,t_{c1})= e^{i\frac{\vec{\lambda}_a \vec{r}_a }{\hbar}} (1-\frac{i}{\hbar} \Delta t H_{ca+\lambda})\psi_a(\vec r_a,t_{c1})$, where $H_{ca+\lambda}=\frac {(\vec p_a +\vec \lambda_a)^2}{2m^*}+V_a$. The demonstration of this result just requires to show that:
\begin{eqnarray}
\label{e5}
&& (\vec p) ^2e^{i\frac{\vec{\lambda}_a \vec{r}_a }{\hbar}}\psi_a(\vec r_a,t)=-\hbar^2 \vec \nabla^2 e^{-i\frac{\vec{\lambda}_a \vec{r}_a }{\hbar}}\psi_a(\vec r_a,t)\\ 
 &&=\vec \lambda_a^2e^{i\frac{\vec{\lambda}_a \vec{r}_a }{\hbar}}\psi_a(\vec r_a,t)-2 i \hbar \vec \lambda_a e^{i\frac{\vec{\lambda}_a \vec{r}_a }{\hbar}}\vec \nabla \psi_a(\vec r_a,t)\nonumber\\ 
 &&-\hbar^2 e^{i\frac{\vec{\lambda}_a \vec{r}_a }{\hbar}} \vec \nabla^2\psi_a(\vec r_a,t)=e^{i\frac{\vec{\lambda}_a \vec{r}_a }{\hbar}} (\vec p +\vec \lambda_a)^2\psi_a(\vec r_a,t).\nonumber
\end{eqnarray}

Therefore, the time evolution of the conditional wave packet at any time $t=t_{c2}+n\Delta t$ can be computed by applying the previous property $n$ times and then:
\begin{eqnarray}
&& (1-\frac{i}{\hbar} \Delta t H_{ca})...(1-\frac{i}{\hbar} \Delta t H_{ca}) e^{i\frac{\vec{\lambda}_a \vec{r}_a}{\hbar}}\psi_a(\vec r_a,t_{c1})\\ 
 &&= e^{i\frac{\vec{\lambda}_a \vec{r}_a }{\hbar}} (1-\frac{i}{\hbar} \Delta t H_{ca+\lambda})... (1-\frac{i}{\hbar} \Delta t H_{ca+\lambda})\psi_a(\vec r_a,t_{c1}).\nonumber
\end{eqnarray}

Finally, we can combine the time evolution of the envelope conditional wave packet $\psi_a(\vec r_a,t)$ before and after the collision in a unique equation of motion as:
\begin{eqnarray}
\label{aschomany}
i \hbar \frac{\partial \psi_a}{\partial t}=\left [\frac{1}{2m^*}\left( \vec p_a +\vec \lambda_a \Theta_{t} \right)^2 +V_a\right] \psi_a
\end{eqnarray}
where $\Theta_{t}(t)$ can be any function which accomplishes that $\Theta_{t}(t)=0$ before the collision ($t<t_{c1}$) and $\Theta_{t}(t)=1$ after the collision ($t>t_{c2}$). For practical purposes and to facilitate computations, in the numerical results we consider $\Theta_{t}\equiv\Theta_{t_c}$ to be the Heaviside step function, $t=t_c$ the time when the interaction occurs and $t_{c2}$ a time infinitely small after $t_{c}$ and $t_{c1}$ a time infinitely small before $t_{c}$. Time interval $t_{c2}-t_{c1}$ can be roughly estimated from time-energy uncertainty relations and it gives a value on the order of few fs. If a more slow/adiabatic evolution of the collision is required in some practical implementations, the equation of motion in \eref{aschomany} can be easily adapted to a slower or more adiabatic collision process by just splitting the whole momentum exchange taking place during one time step of the simulation into more steps but with smaller momentum exchange. This equation of motion of the conditional wave function reproduces \eref{schomany} for the conditional wave packet suffering an electron-lattice interaction with parabolic energy bands. We emphasize that the stochasticity is introduced into \eref{aschomany} because the exact (Bohmian) path of the ions $\vec R^j[t]$ is not explicitly simulated. Their effect is introduced into the dynamics of the electron $\vec r_a^j[t]$ through the random selection of collision times and phonon modes satisfying some \emph{well-known} probability distributions. 

By construction, the time evolution of $\psi(\vec r_a,t)$ before and after the collision is fully coherent. The main and important difference is the change of momentum. For example, in a double barrier, a collision adding and subtracting the momentum $\vec \lambda_a=\hbar \vec q_p$ in the wave function $\psi(\vec r_a,t)$ can convert a non resonant state into a resonant one or vice versa. Until here, only collisions within a unique band have been considered. The implementation in electron-phonon multibands models (already indicated below \eref{g_par}) or other types of collision could be straightforwardly done.

\section{Dirac (linear band) equation}
\label{apf}
In the linear case, $E(\vec p_a)=\pm v_f |\vec p_a|$, with $v_f$ the Fermi velocity. The same development done for the Schr\"odinger equation can be followed here for the evolution of the 2D bispinor solution of the Dirac equation, with a slight difference appearing because the wave function is a bispinor wave function. The Bloch energy eigenstates $| \vec k_a\rangle$ defined in \eref{bloch} have to be substituted by $| \vec k_a,s_a\rangle$ defined as:
\begin{eqnarray}
\phi_{\vec k_a,s_a}=\langle \vec r_a| \vec k_a,s_a \rangle=\dfrac{u_{\vec k_a}e^{i\vec k_a \vec r_a}}{\sqrt{2}} \begin{pmatrix}
1\\
s_a e^{i\beta_{\vec k_a}}
\end{pmatrix},
\label{eigne2Dg}
\end{eqnarray}
where $s_a$ indicates if the electron is in the conduction ($s_a=1$) or valence ($s_a=-1$) band, with positive or negative energies, respectively.  We have defined $e^{i\beta_{\vec k_i}}=\frac{k_{ix}+ik_{iy}}{\sqrt{k_{ix}^2+k_{iy}^2}}$ and $\beta_{\vec k_i}$ the angle of the $\vec k_i$ wave vector. 

All developments done previously for a parabolic band can be reproduced here by just introducing the appropriate  index $s_a$ and the bispinor. In particular, the initial conditional envelope wave packet before the collision in \eref{enbefore} is rewritten here as:
\begin{eqnarray}
&&\psi_a(\vec r_a,t)=\begin{pmatrix}
\psi_{a,1}(\vec r_a,t)\\
\psi_{a,2}(\vec r_a,t)
\end{pmatrix}=\sum_{\vec k_w}\begin{pmatrix}
1\\
s_a e^{i\beta_{\vec{k}_w}}
\end{pmatrix}f_a(\vec{k}_w,t)e^{i\vec{k}_w\vec r_a} \nonumber\\ 
 &&\approx
 \begin{pmatrix}
1\\
s_{0a} e^{i\beta_{\vec k_{0a}}}
\end{pmatrix}\sum_{\vec k_w}f_a(\vec{k}_w,t)e^{i\vec{k}_w\vec r_a},
\label{initialwf}
\end{eqnarray}
where we have assumed again that $\vec k_w \approx \vec k_{0a}$ and $s_a \approx s_{0a}$ with $s_{0a}$ indicating that the initial wave packet belongs to the conduction ($s_{0a}=1$) or valence ($s_{0a}=-1$) band. Identically, the coupling constant defined in \eref{g_par} has to be substituted by the new one: 
\begin{eqnarray}
&&g^{\vec {q}_p}_{\vec k_a,s_a,s_a'}\!=\!\!-i \vec Q_{\vec {q}_p} \; \vec {q}_p \; U_{\vec {q}_p} \int_{\vec {r'}_a} d \vec {r'}_a {e^{-i(\vec k_a+\vec q_p) \vec {r'}_a}}\\ 
 &&u_{\vec k_a+\vec q_p}(\vec {r'}_a) {e^{i\vec k_a \vec {r'}_a}}u_{\vec k_a}(\vec {r'}_a) e^{i\vec v \vec {r'}_a}\left(\frac{1+s_a\; s_a'\;e^{i(\beta_{\vec k_a}-\beta_{\vec k_a+\vec q_p})}}{2}\right),\nonumber
\label{g_lin}
\end{eqnarray}
which contains the information of the transition from the initial energy branch $s_a$ to the final branch $s_a'$. We assume again that, in a particular experiment $j$, only one $\vec q_p[t]$ (or none) is relevant at each time and that $g^{\vec {q}_p,j}_{\vec k_a,s_a,s_a'}[t] \approx g^{\vec {q}_p,j}_{\vec k_0a,s_{0a},s_{fa}}[t]$  where $s_{fa}$ indicates that the final wave packet is in the conduction ($s_{fa}=1$) or valence ($s_{fa}=-1$) band (more exotic collisions with final presence of the wave packet at both energy branches can be considered by two collisions with the one final-branch-collision process mentioned here). Then, the condition given in \eref{after} between the envelope conditional wave packet before and after the collision with parabolic energy bands can be straightforwardly rewritten here as: 
\begin{eqnarray}
&&\psi_a(\vec r_a,t_{c2})=\sum_{\vec k_w} \begin{pmatrix}
1\\
s_{fa} \; e^{i\beta_{\vec{k}_w+\vec q_p}}
\end{pmatrix}f_a(\vec{k}_w,t_{c1})e^{i\vec{k}_w\vec r_a}e^{i\vec q_p\vec r_a}\nonumber\\ 
 &&=\! e^{i\frac{\vec{\lambda}_a \vec{r}_a }{\hbar}}\!\! \sum_{\vec k_w} \begin{pmatrix}
1\\
 s_{fa} s_{0a} e^{i(\beta_{\vec{k}_w+\vec q_p}-\beta_{\vec{k}_w})} s_{0a} e^{i\beta_{\vec{k}_w}}
\end{pmatrix}\!\!f_a(\vec{k}_w,t_{c1})e^{i\vec{k}_w\vec r_a}\nonumber \\ 
&&\approx e^{i\frac{\vec{\lambda}_a \vec{r}_a }{\hbar}}
 \begin{pmatrix}
1 & 0\\
 0 & e^{i \alpha_a}
\end{pmatrix}
\begin{pmatrix}
1\\
s_{0a}e^{i\beta_{\vec{k}_w}}
\end{pmatrix}\sum_{\vec k_w}f_a(\vec{k}_w,t_{c1})e^{i\vec{k}_w\vec r_a},
\label{condi2D}
\end{eqnarray}
where we have introduced  $s_{0a}\; s_{0a}=1$ and $e^{-i\beta_{\vec{k}_w}} e^{i\beta_{\vec{k}_w}}=1$. We define $s_{fa}\; s_{0a}=e^{i{m \pi}}$ where $m\pi$ reflects the changing from one branch to the other due to the absorption/emission of the phonon ($m=1$) or the collision without changing ($m=0$). We have finally defined $\alpha_a=m\pi+\beta_{\vec{k}_{fa}}-\beta_{\vec{k}_{0a}}$. We can then rewrite compactly \eref{condi2D} as: 
\begin{eqnarray}
&&\psi_a(\vec r_a,t_{c2})=\begin{pmatrix}
\psi_{a,1}(\vec r_a,t_{c2})\\
\psi_{a,2}(\vec r_a,t_{c2})
\end{pmatrix}\nonumber\\ 
 && \approx e^{i\frac{\vec{\lambda}_a \vec{r}_a }{\hbar}}\begin{pmatrix}
\psi_{a,1}(\vec r_a,t_{c1})\\
e^{i \alpha_a}\psi_{a,2}(\vec r_a,t_{c1})
\end{pmatrix}.
\label{condi2D2}
\end{eqnarray}

With the same development done for the parabolic band, we know that the time-evolution of $\psi_a(\vec r_a,t)$ before the collision is given by the 2D Dirac equation as:
\begin{eqnarray}
\label{e3bis}
&&i \hbar \frac{\partial \psi_a}{\partial t} = \left(v_f \hat \sigma_x p_{ax}+v_f \hat \sigma_y p_{ay} +  (V_a+A_a+iB_a) \hat {I}\right)\psi_a\nonumber\\ 
 &&= \begin{pmatrix}
V_a+A_a+iB_a  & v_f p_a^-\\
v_f p_a^+ & V_a+A_a+iB_a
\end{pmatrix} \psi_a,
\end{eqnarray}
with $\hat \sigma_x= \begin{pmatrix} 0 & 1 \\ 1 & 0\end{pmatrix}$ and $\hat \sigma_y=\begin{pmatrix} 0 & -i \\ i & 0\end{pmatrix}$ the Pauli matrices, $\hat I=\begin{pmatrix} 1 & 0 \\ 0 & 1\end{pmatrix}$ the identity matrix, $\vec p=\{p_{ax},p_{ay}\}=\{-i \hbar \partial_{x},-i \hbar \partial_{y}\}$  and $p^{\pm}_a\!=\!-i\hbar\partial_{x_a}\pm\hbar\partial_{y_a}$. With the same approximations used in \eref{schomany1} for $A_a$ and $B_a$ based on Ref.  \onlinecite{PRLxavier}, we get the following time evolution operator (propagator) from the initial time $t=t_0$ until a time before the collision $t<t_{c1}$ as $\hat {\mathcal  U_a}(t,t_0)=e^{-\frac{i}{\hbar} \int_{t_{0}}^{t} \hat {\mathcal H}_{ca}(t')dt'}$ with $ H_{ca}=v_f (\hat \sigma_x p_{ax}+\hat \sigma_y p_{ay}) +  V_a \hat I$.  Again we can define the time evolution operator for any time larger than the collision $t>t_{c2}$ as: 
\begin{eqnarray}
&&\hat {\mathcal U_a}(t,t_0)=e^{-\frac{i}{\hbar} \int_{t_{c2}}^{t} \hat {\mathcal H}_{ca}(t')dt'} e^{-\frac{i}{\hbar} \int_{t_{c1}}^{t_{c2}} \hat {\mathcal H}_{epa}(t')dt'} \nonumber\\ 
 &&e^{-\frac{i}{\hbar}
\int_{t_{c1}}^{t_{c2}} \hat {\mathcal H}_{sa}(t')dt'} e^{-\frac{i}{\hbar}  \int_{t_0}^{t_{c1}} \hat {\mathcal H}_{ca}(t')dt'},
\end{eqnarray}
with the interacting Hamiltonian given by $\mathcal H_{epa}=-\hbar\vec{\lambda}_a \vec{r}_a \delta(t-t_c) \hat I$ and $\mathcal H_{sa}=-\hbar \delta(t-t_c) \begin{pmatrix} 1 & 0 \\ 0 & \alpha_a \end{pmatrix}$. with this time-dependent Hamiltonian, it can be easily demonstrated that \eref{condi2D2} is satisfied: 
\begin{eqnarray}
&&\psi_a(\vec r_a,t_{c2})= 
 \begin{pmatrix}
e^{i\frac{\vec{\lambda}_a \vec{r}_a }{\hbar}}\psi_{a,1}(\vec r_a,t_{c1})\\
e^{i\frac{\vec{\lambda}_a \vec{r}_a }{\hbar}}e^{i\alpha_a} \psi_{a,2}(\vec r_a,t_{c1})
\end{pmatrix}\\ 
 &&=e^{-\frac{i}{\hbar} \int_{t_{c1}}^{t_{c2}} \hat {\mathcal H}_{epa}(t')dt'} e^{-\frac{i}{\hbar} \int_{t_{c1}}^{t_{c2}} \hat {\mathcal H}_{sa}(t')dt'} \psi_a(\vec r_a,t_{c1}).\nonumber
\end{eqnarray}
It can also be demonstrated that $(p_a^+)e^{i\frac{\vec{\lambda}_a \vec{r}_a }{\hbar}} \psi_{a,1}(\vec r_a,t)=e^{i\frac{\vec{\lambda}_a \vec{r}_a }{\hbar}}(p_a^++\lambda_a^+)\psi_{a,1}(\vec r_a,t)$. Identically, $(p_a^-)e^{i\frac{\vec{\lambda}_a \vec{r}_a }{\hbar}} \psi_{a,2}(\vec r_a,t)=e^{i\frac{\vec{\lambda}_a \vec{r}_a }{\hbar}}(p_a^-+\lambda_a^-)\psi_{a,2}(\vec r_a,t)$, where we have defined $\lambda_a^{\pm}\!=\!\lambda_{ax}\pm i\lambda_{ay}$. Therefore, we have proved for the bispinor:

\begin{eqnarray}
&&(1-\frac{i}{\hbar} \Delta t \hat {\mathcal H}_{ca})....(1-\frac{i}{\hbar} \Delta t \hat {\mathcal H}_{ca}) e^{i\frac{\vec{\lambda}_a \vec{r}_a }{\hbar}}\psi_a(\vec r_a,t_{c1})\\ 
 &&= e^{i\frac{\vec{\lambda}_a \vec{r}_a }{\hbar}} (1-\frac{i}{\hbar} \Delta t \hat {\mathcal H}_{ca+\lambda})... (1-\frac{i}{\hbar} \Delta t \hat {\mathcal H}_{ca+\lambda})\psi_a(\vec r_a,t_{c1}),\nonumber
\end{eqnarray}

with $\hat H_{ca+\lambda}=v_f (\hat \sigma_x (p_{ax}+\lambda_{ax})+\hat \sigma_y (p_{ay}+\lambda_{ay})) + V_a \hat I$. Notice that we have still not considered the effect of the term $e^{-\frac{i}{\hbar}\int_{t_{c1}}^{t_{c2}} \hat {\mathcal H}_{sa}(t')dt'}$. It includes an angle $e^{i\alpha_a}$ in the second element of the bispinor at time $t=t_{c1}$. As discussed in Appendix C, in most practical applications, for simplicity, we assume an instantaneous collision. A slower or more adiabatic collision process is also easily implementable. Finally,  the global equation of motion of the conditional bispinor that includes all mentioned dynamics and is valid for any time, either before or after the collision, is:
\begin{eqnarray}
i \hbar \frac{\partial \psi_a(\vec r_a,t)}{\partial t}= v_f 
\begin{pmatrix}
V_{a}/v_f & p_a^-+ \lambda_a^{-}\Theta_{tc}\\
(p_a^++ \lambda_a^{+}\Theta_{tc})\chi_{t_c} & V_{a}\chi_{t_c}/v_f
\end{pmatrix} \psi_a(\vec r_a,t).\nonumber \\
\label{e3bisbis}
\end{eqnarray}

As we explained below the term $\chi_{t_c}=exp(i(m\pi+\beta_{\vec k_{fa}}-\beta_{\vec k_{0a}})\Lambda_{t_c})$ projects the general bispinor into positive or negative energy states, and for practical purposes in the numerical results we chose $\Theta_{tc}$ to be the Heaviside step function. This equation of motion exactly reproduces \eref{dirac2d2}. We emphasize again that the stochasticity is introduced into \eref{e3bis} because the exact (Bohmian) path of the ions $\vec R^j[t]$ is not explicitly simulated. Their effect is introduced into the dynamics of the electron $\vec r_a^j[t]$ through the random selection of collision times and phonon modes satisfying some \emph{well-known} probability distributions. 

We achieve the same conclusion as in the Schr\"odinger case: the time evolution of $\psi(\vec r_a,t)$ before and after the collision is fully coherent. For example, as it is shown in the text, if a collision occurs with an initial electron whose direction was not perpendicular to a potential barrier (and therefore will not suffer Klein tunneling) and that collision changes the electron direction appropriately, the electron can experience the full Klein tunneling effect.


\begin{thebibliography}{99}

\bibitem{open} 
H. P. Breuer and F. Petruccione, \textit{Theory of Open Quantum
Systems} (Oxford University Press, Oxford, 2002).

\bibitem{positivity}  
B. Vacchini, Phys. Rev. Lett. {\bf 84} (6), 1374 (2000); J. M. Dominy, A. Shabani, and D. A. Lidar, Quantum Inf Proc. \textbf{15}, 1349 (2016); G. Stefanucci, Y. Pavlyukh, A. M. Uimonen, and R. van Leeuwen, Phys. Rev. B \textbf{90}, 115134 (2014).

\bibitem{Zhen} 
L. Di\'osi, Europhys. Lett. \textbf{30}, 63 (1995); B. Vacchini and K. Hornberger, Physics Reports \textbf{478} 71 (2009); Z. Zhan, E. Colom\'es, and X. Oriols, J Comput Electron \textbf{15}, 1206 (2016). 

\bibitem{caldeira} 
A. O. Caldeira and A. Leggett, Physica (Amsterdam) \textbf{121A}, 587 (1983).

\bibitem{lindblad} 
A. Kossakowski, Rep. Math. Phys. \textbf{3}, 247 (1972); G. Lindblad, Commun. Math. Phys. \textbf{48}, 119 (1976); V. Gorini, K. Kossakowski, and E. Sudarshan, J. Math. Phys. \textbf{17}, 821 (1976).

\bibitem{ferialdi} 
L. Ferialdi, Phys. Rev. Lett. {\bf 116}, 120402 (2016); L. Ferialdi, Phys. Rev. A \textbf{95}, 052109 (2017).

\bibitem{vega} 
I. Vega, and D. Alonso, Rev. Mod. Phys. \textbf{89}, 015001 (2017).

\bibitem{GRW} 
G. C. Ghirardi, A. Rimini, and T. Webber, Phys. Rev. D \textbf{34}, 470 (1986); A. Bassi, K. Lochan, S. Satin, T. P. Singh, and H. Ulbricht, Rev. Mod. Phys. \textbf{85}, 471 (2013).

\bibitem{SSE} 
L. Di\'osi, N. Gisin, and W. T. Strunz, Phys. Rev. A \textbf{58}, 1699 (1998); W.T. Strunz, L. Di\'osi and N. Gisin, Phys. Rev. Lett. \textbf{82}, 1801 (1999); W. T. Strunz, Chemical Physics \textbf{268}, 237 (2001); L. Ferialdi and A. Bassi Phys. Rev Lett. \textbf{108}, 170404 (2012).

\bibitem{Gambetta} 
J. Gambetta and H.M. Wiseman, Phys. Rev. A. \textbf{66}, 012108 (2002); J. Gambetta and H.M. Wiseman, Phys. Rev. A. \textbf{68}, 062104 (2003). 

\bibitem{dioosi}
L. Di\'osi, and L. Ferialdi, Phys. Rev. Lett. \textbf{113}, 200403 (2014).

\bibitem{wiseman}
L. Di\'osi, Phys. Rev. Lett. \textbf{100}, 080401 (2008); H.M. Wiseman and J. M. Gambetta, Phys. Rev. Lett. \textbf{101}, 140401 (2008); L. Di\'osi, Phys. Rev. Lett. \textbf{101}, 149902 (2008). 

\bibitem{Bohm}
D. D\"urr, S. Goldstein, and N. Zangh\`i, J. Stat. Phys. \textbf{67}, 843 (1992); D. Bohm Phys. Rev. \textbf{85}, 166 (1952); D. D\"urr and S. Teufel, \textit{Bohmian Mechanics: The Physics and Mathematics of Quantum Theory} (Springer, Berlin, 2009).

\bibitem{ABM}
X. Oriols and J. Mompart, \textit{Applied Bohmian Mechanics: From Nanoscale Systems to Cosmology} (Pan Stanford Publishing, Singapore, 2011).

\bibitem{durr2}
D. D\"urr, S. Goldstein, R. Tumulka, and N. Zangh\`i, Found. Phys. \textbf{35}, 449 (2005). 

\bibitem{fnt}
The fact that we deal with a CP definition, more than just a positive operator can be trivially demonstrated here, coming from the fact that the same $\psi_a^j(t)$ will be obtained if we enlarge the environment with additional degrees of freedom.

\bibitem{PRLxavier}
X. Oriols, Phys. Rev. Lett. {\bf 98}\penalty0 (6), \penalty0 066803, (2007). T. Norsen, D. Marian, and X. Oriols, Synthese, \textbf{192}, 3125 (2015).

\bibitem{applied}
I. Christov, New J. Phys. \textbf{9}, 70 (2007); R. E. Wyatt, \textit{Quantum Dynamics with Trajectories: Introduction to Quantum Hydrodynamics} (Springer, New York, 2005); R. E. Wyatt, and E. R. Bittner, J. Chem. Phys. \textbf{113} 8898 (2000); Y. Goldfarb, I. Degani, and D. J. Tannor, J. Chem. Phys. \textbf{125} 231103 (2006);  E. Gindensperger, C. Meier, and J. A. Beswick, J. Chem. Phys. \textbf{116} 8 (2002) .

\bibitem{improper}
B. D'Espagnat, \textit{Conceptual Foundations of Quantum Mechanics}, Advanced Book Program (Perseus Books, New York, 1976).

\bibitem{proper}
If we ignore what state is actually present in a closed system (all we know is that several states $|\psi_1\rangle, |\psi_2\rangle,...$ are possible with several probabilities $p_1,p_2,...$), the statistical operator  $\rho=\sum_w p_w |\psi_w\rangle \langle \psi_w|$ is unproblematically defined as a \emph{proper} mixture of states. However, the states $|\psi_1\rangle, |\psi_2\rangle,...$ for an open (sub)system have no ontological meaning within standard quantum theory\cite{improper}. Then, the reduced density matrix is an \emph{improper} mixture of states because the states themselves are ontologically undefined in the standard theory. Our ignorance plays a secondary role in an open system.     

\bibitem{JacoboniBook}
C. Jacoboni and L. Paolo, \textit{The Monte Carlo Method for Semiconductor Device Simulation} (Springer, Berlin, 1989).

\bibitem{RTD}
K. L. Jensen and F. A. Buot, J. Appl. Phys. \textbf{67}, 7602 (1990); W. R. Frensley, Solid State Electron. \textbf{31}, 739 (1988).

\bibitem{rossi1}
R.C. Iotti and F. Rossi, EPL \textbf{112}, 67005 (2015); 

\bibitem{rossi2}
F. Rossi \textit{Theory of Semiconductor Quantum Devices} (Springer-Verlag, Berlin 2011)

\bibitem{electdiss} 
R. G. Lake, G. Klimeck, M. P. Anantram and S. Datta, Phys. Rev. B \textbf{48}, 15132 (1993); S. Datta, Phys. Rev. B \textbf{40}, 5830 (1989); M. B\"{u}ttiker,  Phys. Rev. Lett. \textbf{57} 1761 (1986); S. Datta, \textit{Quantum Transport} (Cambridge University Press, Cambridge, UK, 2005)

\bibitem{BITLLES}
G. Albareda, H. L\'{o}pez, X. Cartoix\`{a}, J.~Su\~{n}\'{e} and X. Oriols, Phys. Rev. B, {\bf 82}, \penalty0 085301 (2010); G. Albareda, J. Su\~{n}\'{e}, and X. Oriols, Phys. Rev. B. {\bf 79}, \penalty0 075315,  (2009); A. Alarc\'on, S. Yaro, X. Cartoix\`{a}, and X. Oriols, J. Phys. Condens Matter \textbf{25}, 325601 (2013);  D. Marian, E. Colom\'es and X. Oriols, J. Phys. Condens. Matter \textbf{27} 245302 (2015); F.L. Traversa \textit{et al.}, IEEE Trans. Electron Dev., {\bf 58}, \penalty0 2104 (2011);  D. Marian, E. Colom\'es, Z. Zhan, and X. Oriols, J. Comp. Elect. {\bf 14}, 114 (2015); A. Alarc\'{o}n and X. Oriols, J. Stat. Mech. \textbf{2009} 01051  (2009). D. Marian, and E. Colom\'{e}s, Fluct. Noise Lett. \textbf{15}, 1640008 (2016).

\end{thebibliography}
\end{document}